\documentclass[12pt]{article}

\usepackage{amssymb, amsmath, amsthm, amsfonts, graphicx, color, bm, float} 
\usepackage[margin=1in]{geometry}
\usepackage{hyperref}
\usepackage{chicago}
\usepackage{setspace} 
\usepackage{graphicx, subfig, epstopdf, enumitem}
\usepackage{caption}

\usepackage{algorithm, algpseudocode}

\newcommand{\bs}[1]{\boldsymbol{#1}}

\newfloat{algorithm}{htbp}{loa}
\floatname{algorithm}{Algorithm}

\DeclareMathOperator*{\argmax}{arg\,max}
\newcommand{\yobs}{{\bs{y}_\mathrm{obs}}}
\newcommand{\sobs}{{\bs{s}_\mathrm{obs}}}
\newcommand{\bftheta}{\bs{\theta}}
\newcommand{\bfy}{\bs{y}}

\newcommand{\bfp}{\bs{p}}
\newcommand{\bfP}{\bs{P}}
\newcommand{\bfs}{\bs{s}}

\title{Recalibration: A post-processing method for approximate Bayesian computation}
\date{\today}

\author{G. S. Rodrigues\footnote{School of Mathematics and Statistics, University of New South Wales, Australia.}\:\,\footnote{CAPES Foundation, Ministry of Education of Brazil, Brazil}\,,
D. Prangle\footnote{School of Mathematics and Statistics, Newcastle University, UK.}\:\:  and S. A. Sisson$^{*}$\footnote{Communicating Author: {\tt Scott.Sisson@unsw.edu.au}}\,}

\begin{document}

\maketitle

\begin{abstract} 
\noindent A new recalibration post-processing method is presented to improve the quality of the posterior approximation when using Approximate Bayesian Computation (ABC) algorithms. Recalibration may 
be used in conjunction with existing post-processing methods, such as regression-adjustments. In addition, this work extends and strengthens the links between ABC and indirect inference algorithms, allowing more extensive use of misspecified auxiliary models in the ABC context. The method is illustrated using simulated examples to demonstrate the effects of recalibration under various conditions, and through an application to an analysis of stereological extremes both with and without the use of auxiliary models. Code to implement recalibration post-processing is available in the {\tt R} package, {\tt abctools}.
\\

\noindent Key words: Approximate Bayesian computation (ABC); Indirect inference; Coverage; Regression-adjustment.
\end{abstract}

\section{Introduction}
\label{sec:introduction}

Approximate Bayesian Computation (ABC) refers to a class of algorithms designed to sample from an approximation to the posterior distribution without directly evaluating the likelihood function. These techniques have expanded the reach of statistical inference to a range of problems where the likelihood function is computationally intractable, in that it is prohibitively expensive or even impossible to evaluate. Instead, inference is based on the ability to simulate data from the model of interest (e.g. \shortciteNP{Beaumont2002,fearnhead2012,sisson+fb17}).

Consider the usual Bayesian setting with a parameter vector $\bftheta=(\theta_1, \theta_2, \ldots, \theta_d)^\top$, a prior $\pi(\bftheta)$, and a model for data $\bfy$, $p(\bfy|\bftheta)$. Let $\yobs$ denote the observed data. In its simplest implementation, ABC repeatedly executes two steps: sampling $(\bftheta, \bfy)$ from the (prior predictive) generative process $\pi(\bftheta) p(\bfy|\bftheta)$, and accepting $\bftheta$ if $\bfy \approx \yobs$ according to some distance measure.
This second step is commonly implemented in an importance sampling framework whereby a weight $w(\bftheta)$ is attached to $\bftheta$ of the form $w(\bftheta)\propto K_h(\|\bfs-\sobs\|)$, where $\bfs=S(\bfy)$ maps $\bfy$ to a low dimensional vector of summary statistics, $\sobs=S(\yobs)$, and $K_h$ is a smoothing kernel with scale parameter $h\geq 0$.
The idealised algorithm where only exact matches $\bfy=\yobs$ are accepted ($h=0$) would produce samples from the exact posterior $\pi(\bftheta|\bfy)$ (or more generally the partial posterior $\pi(\bftheta|\sobs)$, if matching $\bfs=\sobs$).
In practice, approximate matches based on weights $w(\bftheta)$ are retained to side-step the impossibility of exactly matching simulated and observed data in all but the simplest settings.
However this necessity accordingly introduces an approximation error to the ABC posterior approximation. In general, the ABC posterior approximation can be expressed as
\begin{eqnarray}
\label{eqn:abcpostapprox}
	\pi_{ABC}(\bftheta|\sobs) = \int K_h(\|S(\bfy)-\sobs\|)p(\bfy|\bftheta)\pi(\bftheta)d\bfy.
\end{eqnarray}
See e.g. \shortciteN{sisson+fb17} for further details.

A number of post-processing techniques have been proposed to correct this approximation error once samples from the ABC posterior approximation have been obtained, resulting in an estimate $\hat{\pi}_{ABC}(\bftheta|\sobs)$ which better approximates the true (partial) posterior $\pi(\bftheta|\sobs)$ than (\ref{eqn:abcpostapprox}). \shortciteN{Beaumont2002} introduced a regression-adjustment approach, in which the ABC samples are corrected with the aid of a local linear regression model for $\bftheta|\bfs-\sobs$, fitted to  the $(\bftheta, \bfs)$ samples from (\ref{eqn:abcpostapprox}). 
Various extensions to this technique  include non-linear, heteroscedastic regression \cite{Blum2010}, and ridge regression adjustments \shortcite{blum+nps13}. However, there is some evidence emerging to suggest that regression-adjustments tend to overcorrect and produce approximate posteriors that are too precise, leading to nominal credible intervals with coverage much higher than should occur under $\pi(\bftheta|\sobs)$ \shortcite{marin+rprr16,frazier+rr17}.
From the perspective of marginal density estimation,  \shortciteN{nott+fms14} (see also \shortciteNP{li+nfs17}) developed a marginal-adjustment which replaces  low-dimensional marginal distributions of (\ref{eqn:abcpostapprox}) by more accurate marginal distributions estimated using smaller numbers of summary statistics than in $\bfs$. This exploits the fact that ABC methods are known to perform poorly for larger numbers of summary statistics due to the curse of dimensionality in the comparison $\|\bfs-\sobs\|$, however this approach requires the identification of subsets of summary statistics that are informative for each margin, which may not be easily available.

In this paper we introduce a novel \emph{recalibration} post-processing method for improving the accuracy of the ABC posterior approximation that avoids the problems of existing post-processing techniques. It is based on the ideas in 
\shortciteN{Prangle2013}, who derive a diagnostic tool for ABC based on the so-called \emph{coverage property} \shortcite{Cook2006,fearnhead2012,Prangle2013},
which tests whether for a given $h>0$ the estimated marginals of $\pi_{ABC}(\bftheta|\sobs)$ (or $\hat{\pi}_{ABC}(\bftheta|\sobs)$)  are well ``calibrated''.
Calibration requires that  estimated credible intervals have the correct probabilities of containing the true parameter values.
If calibration does not hold, \shortciteN{Prangle2013} suggest reducing $h$ until it does hold. However, this is not always feasible, particularly as reducing $h$ increases the Monte Carlo error of the Monte Carlo sample approximation of (\ref{eqn:abcpostapprox}) for a fixed computational budget.

Our approach extends the ideas in \shortciteN{Prangle2013} to develop a post-processing recalibration adjustment that aims to produce an
approximation $\hat{\pi}_{ABC}(\bftheta|\sobs)$ that is  well calibrated.
Our method achieves this approximately and,  as a result, the coverage problems associated with the regression adjustment \shortcite{marin+rprr16} can be mitigated
by construction. Recalibration can be applied directly to samples from $\pi_{ABC}(\bftheta|\sobs)$, or to improve the output from other post-processing adjustments. 
Recalibration is related to indirect inference -- a technique in which  inference is performed with the aid of an auxiliary misspecified model \shortcite{Gourieroux2016}. The use of indirect inference in the ABC framework has been previously explored by \shortciteN{Drovandi2015}, \shortciteN{drovandi+mr17}. Our approach also relates to procedures that correct the biases in an initial estimate based on simulation under the model \shortcite{Menendez2014}.

We introduce our recalibration approach in Section \ref{sec:recalibration}.
We demonstrate its performance in two simulation studies  in Section \ref{sec:simulationstudies}, using a Gaussian auxiliary posterior estimator for inference on a sum of lognormals distribution, and a standard ABC analysis of a ``twisted normal'' model.
Section \ref{sec:application} revisits the analysis of \citeN{Erhardt} in a real stereological extremes problem and shows that the recalibration adjustment can correct the bias of their regression-adjustment ABC implementation. We conclude with a discussion of the merits and limitations of recalibration in Section \ref{sec:discussion}, including the possibility of correcting approximate Bayesian inference methods beyond ABC.

\section{Recalibration}
\label{sec:recalibration}

\subsection{Motivation}
\label{sec:motivation}

Our recalibration post-processing procedure is based on the {\em coverage} property. 
An $\alpha\%$ credible region for a parameter $\bftheta$ is a region $R$ with the property that $\mbox{Pr}(\bftheta\in R|\yobs)=\alpha/100$. Loosely, the coverage property asserts that for data $\bfy_0$ generated under the model for a known parameter value $\bftheta_0=(\theta_{0,1},\ldots,\theta_{0,d})^\top$, so that $\bfy_0\sim p(\bfy|\bftheta_0)$, credible intervals constructed from the posterior $\pi(\theta|\bfy_0)$ will have the claimed probability of containing $\bftheta_0$.

Coverage has been previously examined in the ABC literature. Most commonly it has been used to validate analyses (e.g. \shortciteNP{Wegmann2009,Wegmann2010,Aeschbacher2012}), with \shortciteN{Prangle2013} extending coverage ideas to develop testable diagnostics to determine whether the marginals of $\pi_{ABC}(\bftheta|\sobs)$ are different to those of $\pi(\bftheta|\sobs)$, and similarly whether estimated model probabilities under ABC are different to the true posterior model probabilities given $\sobs$ in a multi-model analysis. Coverage is identified as a desirable property of ABC posterior distributions by \citeN{fearnhead2012}, who also introduce `noisy ABC' which automatically satisfies the coverage property, and \shortciteN{Menendez2014} use related ideas to correct bias in ABC credible intervals. Finally, the failure of regression adjustment techniques to produce ABC approximations $\hat{\pi}(\theta|\sobs)$ that satisfy the coverage property, is being used as evidence that they are producing poor approximations \shortcite{marin+rprr16,frazier+rr17}.

Our recalibration adjustment is closely linked to the diagnostic techniques of \shortciteN{Prangle2013}. 
Let $F_{\bfs}(\bftheta)$ be the distribution function of $\pi(\bftheta|\bfs)$, the partial posterior for $\bftheta$ given some summary dataset $\bfs$, and $F_{j,\bfs}(\theta_j)$ be the $j$-th associated marginal distribution function, for $j=1,\ldots,d$. Our interest is sampling from $F_\sobs(\bftheta)$, the partial posterior distribution given the observed data summary  $\sobs$.

For some choice of parameter $\bftheta_0$, and generated dataset $\bfs_0=S(\bfy_0)$ with $\bfy_0\sim p(\bfy|\bftheta_0)$,  \shortciteN{Prangle2013} demonstrated that the location of the $j$-th marginal parameter $\theta_{0,j}$ in the $j$-th marginal posterior distribution of $\pi(\bftheta|\bfs_0)$, as measured by $p_j=F_{j, \bfs_0}(\theta_{0,j}):=\mbox{Pr}(\theta_j<\theta_{0,j}|\bfs_0)$ will give $p_j\sim U(0,1)$ for $j=1,\ldots,d$.
This then allows for the basis of a test for whether $\tilde{F}_{j,\sobs}(\theta_j)$, the $j$-th marginal distribution function of the ABC posterior approximation $\pi_{ABC}(\bftheta|\sobs)$, is the same as the true marginal distribution function, i.e. whether $\tilde{F}_{j,\sobs}(\theta_j)=F_{j,\sobs}(\theta_j)$.

This test proceeds by generating $(\bftheta^{(i)},s^{(i)})$ pairs, $i=1,\ldots,N$, from $\bftheta^{(i)}\sim\pi(\bftheta)$ (or other suitable distribution) and $\bfs^{(i)}=S(\bfy^{(i)})$, $\bfy^{(i)}\sim p(\bfy|\bftheta^{(i)})$, and constructing the ABC posterior approximation $\pi_{ABC}(\bftheta|\bfs^{(i)})$ for each $\bfs^{(i)}\in A(\sobs)$, where $A(\sobs)$ is some set centred around $\sobs$. 
Then, for each $\bfs^{(i)}\in A(\sobs)$, the statistics $p_j^{(1)},\ldots,p_j^{(N)}$, where $p_j^{(i)}=\tilde{F}_{j,\bfs^{(i)}}(\theta_j^{(i)})$, will only be distributed as $U(0,1)$ if $\tilde{F}_{j,\bfs^{(i)}}(\theta_j)=F_{j,\bfs^{(i)}}(\theta_j)$, which can be determined via standard tests of uniformity for each margin $j=1,\ldots,d$. If this test is satisfied, then it can be inferred that the marginal distributions of $\tilde{F}_{\sobs}(\bftheta)$ are approximately those of $F_{\sobs}(\bftheta)$ and that, marginally at least, the ABC posterior approximation $\pi_{ABC}(\bftheta|\sobs)$ is a good approximation of $\pi(\bftheta|\sobs)$.
(Note that in practice, $\pi_{ABC}(\bftheta|\bfs)$ and $\tilde{F}_{j,\bfs}$ are constructed from weighted samples.)

We now extend this idea. However, rather than merely testing whether there are significant marginal deviations between $\tilde{F}_{\sobs}(\bftheta)$ and $F_\sobs(\bftheta)$, we use the measured differences to adjust those samples $\bftheta$ from $\pi_{ABC}(\bftheta|\sobs)$ so that $\hat{\tilde{F}}_{j,\sobs}(\bftheta)\approx F_{j,\sobs}(\bftheta)$ is a good approximation (where $\hat{\tilde{F}}_{j,\sobs}(\bftheta)$ is the $j$-th marginal distribution function of the adjusted samples). That is, that the resulting post-processed approximation $\hat{\pi}(\bftheta|\sobs)$, approximately satisfies the coverage property, and is accordingly approximately well calibrated.

\subsection{Method}

So far we have assumed that $\tilde{F}_{j,\bfs}(\theta_j)$, the $j$-th marginal distribution of $\tilde{F}_{\bfs}(\bftheta)$, is the $j$-th marginal distribution function of the ABC posterior approximation $\pi_{ABC}(\bftheta|\bfs)$. However, all that is required to implement the recalibration adjustment is that some approximate method for inferring the posterior marginal distribution functions is available. Such approximate methods arise from adopting 
auxiliary models which approximate $\pi(\bftheta|\bfs)$ with different posterior forms, such as those obtained under
the Bayesian indirect inference framework \shortcite{drovandi+mr17,Drovandi2015},
variational Bayes \shortcite{tran+nk17}, regression density estimation \shortcite{fan+ns13} and 
expectation-propagation (exponential family) based approximations \cite{barthelme+c14}.
We now suppose that  $\tilde{F}_{\bfs}(\bftheta)$ 
and the associated marginal distribution functions $\tilde{F}_{j,\bfs}(\theta_j)$, $j=1,\ldots,d$, are available as approximations to $F_{\bfs}(\bftheta)$ and $F_{j,\bfs}(\theta_j)$, based on some auxiliary model, which may include the standard ABC posterior approximation $\pi_{ABC}(\bftheta|s)$. Note that the recalibration adjustment will only make use of the marginal distribution functions $\tilde{F}_{j,\bfs}(\theta_j)$, and not the joint distribution function $\tilde{F}_{\bfs}(\bftheta)$,
and that these approximate marginal distribution functions are assumed to have a well defined inverse, $\tilde{F}^{-1}_{j,\bfs}(\cdot)$.

In order to state the recalibration adjustment, first define
\[
G_{\bfs}(\bfp) = F_{\bfs}[(\tilde{F}_{1,\bfs}^{-1}(p_1),\ldots, \tilde{F}_{d,\bfs}^{-1}(p_d))^\top]
\]
where $\bfp=(p_1,p_2,\ldots,p_d)^\top \in [0,1]^d$ (where $d$ is the number of parameters). The function $G_{\bfs}(\bfp)$ 
incorporates the posterior dependence structure of $\pi(\bftheta|\bfs)$, through $F_{\bfs}(\cdot)$, but it also provides a connection between the true (through $F_{\bfs}(\bftheta)$) and the estimated marginal posterior quantile functions $\tilde{F}_{j,\bfs}^{-1}(p_j)$.
We now provide several simple results on $G_{\bfs}(\bfp)$ which will be useful to establish the recalibration adjustment.

\paragraph{Result 1} 
Suppose a random variable $\bfP = (P_1, \ldots,P_d)^\top$ has distribution $G_{\bfs}(\bfp)$.
Then $P_j | \bfs \sim U(0,1)$ for $j=1,\ldots,d$, if and only if the estimated marginal posteriors $\tilde{F}_{j,\bfs}(\cdot)$ equal the true marginal posteriors $F_{j,\bfs}(\cdot)$.

\begin{proof}
First suppose that $\tilde{F}_{j,\bfs}(\cdot) = F_{j,\bfs}(\cdot)$.
Then the $j$-th marginal distribution function of $G_{\bfs}(\bfp)$ is $F_{j,\bfs}[\tilde{F}_{j,\bfs}^{-1}(p_j)] = p_j$, which is a
$U(0,1)$ distribution.
Next suppose that the $j$-th marginal distribution of $G_{\bfs}(\bfp)$ is a $U(0,1)$ distribution.
Then $F_{j,\bfs}[\tilde{F}_{j,\bfs}^{-1}(p_j)] = p_j$.
Let $q_j=\tilde{F}^{-1}_{j,\bfs}(p_j)$. 
Then we have
$F_{j,\bfs}(q_j)=\tilde{F}_{j,\bfs}(q_j)$ as required.
\end{proof}

Result 1 states that $\bfP \sim G_{\bfs}(\bfp)$ is \emph{marginally} uniform if and only if $\tilde{F}_{j,\bfs}(\cdot)=F_{j,\bfs}(\cdot)$, for $j=1,\ldots, d$, but does not comment on its dependence structure.
\shortciteN{Prangle2013} exploited a variant of this result to test whether the marginal distributions of $\pi_{ABC}(\bftheta|\sobs)$ were equal to those of $\pi(\bftheta|\sobs)$ by testing for uniformity of realised $P_i$ values, as described in Section \ref{sec:motivation}.

\paragraph{Result 2}
Suppose that the random variable $\bfP=(P_1,\ldots, P_d)^\top$ has distribution function $G_{\bfs}(\bfp)$.
Then conditional on $\bfs$, $(\tilde{F}_{1,\bfs}^{-1}(P_1), \ldots, \tilde{F}_{d,\bfs}^{-1}(P_d))^\top$ has distribution $F_{\bfs}(\bftheta)$.
\begin{proof}
\begin{align*}
&& 
\Pr(P_1 \leq p_1, \ldots,P_d \leq p_d | \bfs) &= F_{\bfs}[(\tilde{F}_{1,\bfs}^{-1}(p_1), \ldots,\tilde{F}_{d,\bfs}^{-1}(p_d))^\top] \\
\Rightarrow && \Pr(P_1 \leq \tilde{F}_{1,\bfs}(\theta_1),\ldots, P_d \leq \tilde{F}_{d,\bfs}(\theta_d)| \bfs) &= F_{\bfs}((\theta_1, \ldots,\theta_d)^\top) \\
\Rightarrow &&\Pr(\tilde{F}_{1,\bfs}^{-1}(P_1) \leq \theta_1, \ldots,\tilde{F}_{d,\bfs}^{-1}(P_d) \leq \theta_d | \bfs) &= F_{\bfs}(\bftheta)
\end{align*}
as required.
\end{proof}

Result 2 provides  a straightforward way to use an observation from $G_{\bfs}(\bfp)$ to generate a sample from $F_{\bfs}(\bftheta)$. Result 3 below provides the converse -- a way to use an observation from $F_{\bfs}(\bftheta)$ to generate a sample from $G_{\bfs}(\bfp)$.

\paragraph{Result 3}
Suppose that the random variable $\bftheta=(\theta_1,\ldots, \theta_d)^\top$ has distribution function $F_{\bfs}(\bftheta)$.
Then conditional on $\bfs$, $(\tilde{F}_{1,\bfs}(\theta_1), \ldots, \tilde{F}_{d,\bfs}(\theta_d))^\top$ has distribution $G_{\bfs}(\bfp)$.
\begin{proof}
\begin{eqnarray*}
\Pr(\tilde{F}_{1,\bfs}(\theta_1) \leq p_1, \ldots,\tilde{F}_{d,\bfs}(\theta_d) \leq p_d | \bfs)
 & = & \Pr(\theta_1 \leq \tilde{F}_{1,\bfs}^{-1}(p_1),\ldots, \theta_d \leq \tilde{F}_{d,\bfs}^{-1}(p_d) | \bfs) \\
 & = & F_{\bfs}[(\tilde{F}_{1,\bfs}^{-1}(p_1), \ldots,\tilde{F}_{d,\bfs}^{-1}(p_d))^\top]
\end{eqnarray*}
as required.
\end{proof}

These results may be combined in a procedure to recalibrate the ABC posterior approximation. For simplicity of presentation, we first focus on the recalibration of samples drawn from $\pi_{ABC}(\bftheta|\sobs)$ (or $\hat{\pi}_{ABC}(\bftheta|\sobs)$) under the standard ABC implementation. Following this, in Section \ref{sec:aux} we describe how recalibration can also  be implemented using an auxiliary estimator.

A standard ABC posterior simulation algorithm, complete with the recalibration procedure,  is outlined in Algorithm  \ref{alg:recalibration}. More sophisticated versions of ABC algorithms could be used.
In Algorithm \ref{alg:recalibration}, simulation from $\pi_{ABC}(\bftheta|\sobs)$  begins by drawing $N$ parameter and summary statistic pairs $\{(\bftheta^{(i)},\bfs^{(i)})\}_{i=1}^N$ from $\bftheta^{(i)}\sim\pi(\bftheta)$ and $\bfs^{(i)}=S(\bfy^{(i)})$ where $\bfy^{(i)}\sim p(\bfy|\bftheta^{(i)}$). These samples are then used to approximate $\pi(\bftheta|\sobs)$ by weighting them by $w^{(i)}\propto K_h(\|\bfs^{(i)}-\sobs\|)$.
From this posterior approximation, the marginal distribution functions $\tilde{F}_{j,\sobs}(\theta_j)$ based on $\sobs$ can be constructed by e.g. the empirical cdf or by smoothed versions of such.

\begin{algorithm}[tb] 
\caption{Recalibration of ABC output}
\label{alg:recalibration}
   
  \noindent {\it Inputs:}
  \begin{itemize}[noitemsep]
  \item An observed dataset $\yobs$.
  \item A prior $\pi(\bftheta)$ and intractable generative model $p(\bfy|\bftheta)$, with $\bftheta=(\theta_1,\ldots,\theta_d)^\top$.
  \item An observed vector of summary statistics $\sobs=S(\yobs)$.
  \item A smoothing kernel $K_h(u)$ with scale parameter $h>0$.
  \item A positive integer $N$ defining the number of ABC samples.
  \end{itemize}

  \noindent {\it Data simulation and weighting:}
  
  \noindent For $i=1, \ldots, N$: 
  \begin{enumerate}[noitemsep]
  \item[1.1] Generate $\bftheta^{(i)} \sim \pi(\bftheta)$ from the prior. 
  \item[1.2] Generate $\bs{y}^{(i)} \sim p(\bfy|\bftheta^{(i)})$ from the likelihood. 
  \item[1.3] Compute the summary statistics $\bfs^{(i)}=S(\bfy^{(i)})$.  
  \item[1.4] Compute the sample weight $w^{(i)} \propto K_h(||\bfs^{(i)}-\sobs||)$.
  \end{enumerate}
  
  \noindent {\it Recalibration:}
  \begin{enumerate}
    \item[2.1] For $j=1,\ldots,d$, construct $\tilde{F}_{j,\sobs}(\cdot)$ based on the samples $\{(\bftheta^{(i)},w^{(i)})\}_{i=1}^N$.
  \end{enumerate}
For each $i$ such that $w^{(i)}>0$,     and for $j=1, \ldots, d$:
  \noindent   \begin{enumerate}[noitemsep]
    \item[2.2] Construct $\tilde{F}_{j,\bfs^{(i)}}(\cdot)$ based on the samples $\{(\bftheta^{(k)},\bfs^{(k)})\}_{k=1,k\neq i}^N$ using the same procedure as in steps 1.4 and 2.1.

\item[2.3] Set $p^{(i)}_j= \tilde{F}_{j, \bfs^{(i)}}(\theta^{(i)}_j)$.
     \item[2.4] [Optional] Correct $p_j^{(i)}$ using a regression-adjustment (see Section \ref{sec:regadj}).
    \item[2.5] Set $\hat{\theta}^{(i)}_j=\tilde{F}^{-1}_{j, \sobs} (p^{(i)}_j)$.
  \end{enumerate}
  
  \noindent {\it Outputs:} 
  \begin{itemize}
  \item Standard ABC output: a set of weighted samples $\{(\bftheta^{(i)}, w^{(i)})\}_{i=1}^N$ from $\pi_{ABC}(\bftheta|\sobs)$.
  \item A set of recalibrated weighted samples $\{(\hat{\bftheta}^{(i)}, w^{(i)})\}_{i=1}^N$ from the recalibrated approximate posterior $\hat{\pi}_{ABC}(\bftheta|\sobs)$.
\end{itemize}
\end{algorithm}

For each of these (weighted) samples $\bftheta^{(i)}|w^{(i)}>0$ used, 
an individual recalibration adjustment is performed.
Firstly, samples are first drawn from the ABC posterior $\pi_{ABC}(\bftheta|\bfs^{(i)})$ in the same manner as for those drawn from $\pi_{ABC}(\bftheta|\sobs)$.
It is possible to avoid the cost of performing a full ABC analysis by reusing the simulations from steps 1.1--1.3 of Algorithm \ref{alg:recalibration}, as is relatively common for ABC algorithms \shortcite{blum+nps13,Prangle2013}.
From the samples from $\pi_{ABC}(\bftheta|\bfs^{(i)})$, the marginal distribution functions $\tilde{F}_{j,\bfs^{(i)}}(\cdot)$ can be constructed, for $j=1,\ldots,d$, and the corresponding vector $\bfp^{(i)}=(p_1^{(i)},\ldots,p_d^{(i)})^\top$ obtained via   $p_j^{(i)}=\tilde{F}_{j,\bfs^{(i)}}(\theta_j^{(i)})$. Since $\bftheta^{(i)}$ is an exact draw from the posterior distribution $\pi(\bftheta|\bfs^{(i)})$, then Result 3 states that $\bfp^{(i)}$ is an exact draw from $G_{\bfs^{(i)}}(\bfp)$.

If the ABC  method produces the exact posterior so that $\pi_{ABC}(\bftheta|\bfs^{(i)})=\pi(\bftheta|\bfs^{(i)})$, 
then Result 1 (see also \shortciteNP{Prangle2013}) states that the resulting marginal distributions of $p_j^{(i)}$ would be $U(0,1)$. Of course, this is unlikely to be the case in practice, and so the marginal distributions $\tilde{F}_{j,\bfs^{(i)}}$ characterise the deviations away from uniformity, such as bias, or over-/under-estimation of variance.
These deviations, contained within the marginal $p_j^{(i)}$, are then mapped onto the quantiles of the original ABC approximation of $\pi(\bftheta|\sobs)$, producing the adjusted sample $\hat{\bftheta}^{(i)}=(\hat{\theta}_1^{(i)},\ldots,\hat{\theta}_d^{(i)})^\top$ where $\hat{\theta}_j^{(i)}=\tilde{F}^{-1}_{j,\sobs}(p_j^{(i)})$ for $j=1,\ldots,d$.

If $G_{\bfs^{(i)}}(\bfp) = G_\sobs(\bfp)$, then Result 2 states that the resulting $\hat{\bftheta}^{(i)}$ would be a draw from $F_{\sobs}(\bftheta)$, the exact (partial) posterior.
In practice, however, it must be assumed that  $G_{\bfs^{(i)}}(\bfp) \approx G_\sobs(\bfp)$, and so the recalibrated draws $\hat{\bftheta}^{(i)}$ will be draws from an approximation to $F_{\sobs}(\bftheta)$. However, if similar biases and deviations away from the true posterior based on the approximation of $\pi(\bftheta|\bfs^{(i)})$ are similar to those present in the approximation of $\pi(\bftheta|\sobs)$, then the recalibration of an exact sample $\bftheta^{(i)}$ from $\pi(\bftheta|\bfs^{(i)})$ to $\hat{\bftheta}^{(i)}$ approximately from $\pi(\bftheta|\sobs)$ can be expected to be beneficial.
We explore how well this works in practice in Section \ref{sec:simulationstudies}.

\subsection{Recalibration with an auxiliary estimator}
\label{sec:aux}

Algorithm \ref{alg:recalibration} recalibrates the weighted samples $\{(\bftheta^{(i)},w^{(i)})\}_{i=1}^N$ from steps 1.1--1.4 by constructing a model to approximate the posterior distribution $\pi(\bftheta|\bfs)$ -- namely $\pi_{ABC}(\bftheta|\bfs)$ -- and construct the univariate marginals $\tilde{F}_{j,\bfs}(\cdot)$ required for the recalibration. However the ABC posterior $\pi_{ABC}(\bftheta|\bfs)$ is not the only model that can be used for this task.

Suppose that, more generally, we have an auxiliary model $g(\bfy|\bftheta)$ with an easily computable maximum likelihood estimator $\bfs=S(\bfy)$, so that $g(\bfy|\bftheta)=g(\bfs|\bftheta)$. Motivated by arguments in indirect inference \shortcite{Gourieroux2016,Gleim2013} and Bayesian indirect inference \shortcite{drovandi+mr17,Drovandi2015} the auxiliary model is commonly a close, but tractable surrogate of the intractable model $p(\bfy|\bftheta$). Suppose also that given the prior distribution $\pi(\bftheta)$ it is computationally convenient to fit the associated posterior distribution $g(\bftheta|\bfs)\propto g(\bfs|\bftheta)\pi(\bftheta)$ to $\bfs$. 
In this setting, the univariate marginal distributions of $g(\bftheta|\bfs^{(i)})$ can be constructed as $\tilde{F}_{j,\bfs^{(i)}}(\cdot)$, and subsequently used for the recalibration of the weighted sample $(\bftheta^{(i)},w^{(i)})$ as before. With good choice of $g(\bftheta|\bfs)$ this procedure can be considerably faster and more efficient than using the ABC approximate posterior $\pi_{ABC}(\bftheta|\bfs)$ as the auxiliary estimator.

This use of the auxiliary model is different to some previous usages where the MAP or MLE of the auxiliary model  defined summary statistics that were then used for a standard ABC analysis (e.g. \shortciteNP{Gleim2013,Drovandi2015,Martin2017}). Here, the whole auxiliary model is used to approximate the intractable posterior and produce univariate marginal distributions, rather than merely define a point estimate of the parameters.

Algorithm \ref{alg:recalibration-aux} lists the modifications to Algorithm \ref{alg:recalibration} when using a more general auxiliary model. We explore the use of non-ABC auxiliary models in the simulation study in Section \ref{sec:example}, and directly contrast ABC with non-ABC auxiliary models in the recalibration of an analysis of stereological extremes in Section \ref{sec:application}.

\begin{algorithm}[tb] 
\caption{Recalibration of an auxiliary estimator  (Modifications to Algorithm \ref{alg:recalibration})}
\label{alg:recalibration-aux}
   
  \noindent {\it Inputs:}
  \begin{itemize}[noitemsep]
    \item A tractable auxiliary model for the posterior $\pi(\bftheta|\bfy)$ with accessible maximum likelihood estimate (MLE) $\bfs=S(\bfy)$ that admits auxiliary univariate marginal distribution functions $\tilde{F}_{j,\bfs}(\theta_j)$, $j=1,\ldots,d$. 
  \end{itemize}

  \noindent {\it Data simulation and weighting:}
  
  \noindent For $i=1, \ldots, N$: 
  \begin{enumerate}[noitemsep]
  \item[1.3] Compute the MLE of the auxiliary model $\bfs^{(i)}=S(\bfy^{(i)})$.  
  \end{enumerate}
  
  \noindent {\it Recalibration:}
  \begin{enumerate}
    \item[2.1] For $j=1,\ldots,d$, construct $\tilde{F}_{j,\sobs}(\cdot)$ based on the auxiliary MLE $\sobs$.
  \end{enumerate}
For each $i$ such that $w^{(i)}>0$,     and for $j=1, \ldots, d$:
  \noindent   \begin{enumerate}[noitemsep]
    \item[2.2] Construct $\tilde{F}_{j,\bfs^{(i)}}(\cdot)$ based on the auxiliary MLE $\bfs^{(i)}$.
  \end{enumerate}
  
  \noindent {\it Outputs:} 
  \begin{itemize}
  \item A set of recalibrated weighted samples $\{(\hat{\bftheta}^{(i)}, w^{(i)})\}_{i=1}^N$ approximately from the  posterior $\pi(\bftheta|\sobs)$.
\end{itemize}
\end{algorithm}

\subsection{Regression-adjusted recalibration}
\label{sec:regadj}

There are two natural ways in which regression-adjustment methods can be combined with recalibration in an ABC analysis. The most straightforward is where recalibration is employed to approximately correct for any biases incurred in a standard regression-adjustment ABC analysis (c.f. \shortciteNP{marin+rprr16,frazier+rr17}).

An alternative use of regression adjustment methods stems from the fact that the quality of a recalibrated posterior approximation rests on how well $G_{\bfs^{(i)}}(\bfp)$ approximates $G_{\sobs}(\bfp)$. In the case where there are reasonable differences between $G_{\bfs^{(i)}}(\bfp)$ and $G_{\sobs}(\bfp)$, one approach is to adjust the values of $\bfp^{(i)}$ given the predictors $\bfs^{(i)}$. In the case of a weighted local-linear regression (e.g. \shortciteNP{Beaumont2002}) the model would be
\[
	\eta(\bfp^{(i)}) = {\boldsymbol \alpha} + {\boldsymbol \beta}(\bfs^{(i)}-\sobs) + {\boldsymbol \epsilon}^{(i)}
\]
for $i=1,\ldots,N$, where $\boldsymbol{\alpha}\in\mathbb{R}^d$, $\boldsymbol{\beta}$ is a $d\times \dim(s^{(i)})$ matrix, $\boldsymbol{\epsilon}^{(i)}\sim N_d(0,\Sigma)$,
 $\eta(\cdot)$ is the logistic link function, and where the pair $(\bfp^{(i)},\bfs^{(i)})$ is given the weight $K_h(\|\bfs^{(i)}-\sobs\|)$.
In this manner, the aim is to 
transform $\bfp^{(i)}$ so that if behaves as an approximate sample from $G_{\sobs}(\bfp)$ rather than an exact sample from $G_{s^{(i)}}(\bfp)$. 
Of course for this adjustment to be beneficial it requires that the fitted regression model be highly accurate. If the model is  poorly specified, as with standard regression-adjusted analyses, the final estimation error could easily increase compared to if it is not used. 
Both alternative uses of regression-adjustment with recalibration are examined in
Section \ref{sec:simulation}.

\section{Simulation studies}
\label{sec:simulationstudies}

We now examine the performance of the recalibration procedure of the previous Section on two simulated examples. 
The first makes use of a tractable Gaussian auxiliary model estimator for inference on a sum of lognormals distribution.
The second examines the effect of recalibration on a ``twisted normal'' model under varied ABC inference configurations.

\subsection{A sum of log-normals model}
\label{sec:example}

Consider a univariate random variable $Y = \sum^L_{\ell=1} X_{\ell}$, where $X_{\ell} \sim \mbox{LogNormal}(\mu, \sigma)$ 
are independent and identically distributed log-normal random variables with parameter $\bftheta=(\mu, \sigma)^\top$. 
Log-normal distributions are commonly used to model heavy-tailed quantities, including stock prices and insurance claims. In these settings, $Y$ can represent the complete value of a stock portfolio, or the total liability of claims for an insurance company (particularly if $L$ is also random). 
Despite its structural simplicity, the associated likelihood function $p(\bfy|\bftheta)$ cannot be computed exactly, even numerically, for $L>3$ (For $L=2$ and possibly $L=3$, the likelihood may viably be computed numerically through convolution integrals.) Several methods have been proposed to approximate this function \shortcite{Fenton1960,Schwartz1982,Jingxian2005}, with the Fenton-Wilkinson approximation perhaps the most widely known \shortcite{Fenton1960,asmussen+r08}. Here, the intractable likelihood is approximated by another log-normal distribution with matching first and second moments. More precisely, it is assumed that $p_Y(\bfy|\bftheta) \approx p_Z(\bfy|\bftheta)$, where $Z \sim \mbox{LogNormal}(\alpha, \beta^2)$, with 
\begin{align*}
\alpha  &= \mu + \log L + 0.5(\sigma^2 - \beta^2), \\
\beta^2 &= \log[(\exp(\sigma^2)-1)/L + 1].
\end{align*}

Suppose that we have $n$ observations of $Y$, $\yobs=(y_{\mathrm{obs}, 1}, \ldots, y_{\mathrm{obs}, n})^\top$, and $\pi(\bftheta)$ is defined through the independent marginal prior distributions $\mu \sim N(0, 1)$ and $\sigma^2 \sim \text{Gamma}(1, 1)$, where $\bftheta=(\mu,\sigma)^\top$. While the target posterior $\pi(\bftheta | \bfy)=\pi_Y(\bftheta|\bfy) \propto p_Y(\bfy|\bftheta) \pi(\bftheta)$ is intractable, the approximation $\pi_Z(\theta|\bfy)\propto p_Z(\bfy|\bftheta)\pi(\bftheta)$ is amenable to posterior simulation algorithms such as MCMC. In principle then, this lognormal approximation $\pi_Z(\bftheta|\bfy)$ could be used as the auxiliary posterior model $g(\bftheta|\bfs)$, where $\bfs$ is the MLE of $p_Z(\bfy|\bftheta)$. However, to do this would then require that a posterior simulation algorithm be implemented to draw samples from $\pi_Z(\bftheta|\bfy^{(i)})=g(\bftheta|\bfs^{(i)})$, for each $i$ for which $w^{(i)}=K_h(\|\bfs^{(i)}-\sobs\|)>0$, in order to construct the $\tilde{F}_{j,\bfs^{(i)}}(\cdot)$ marginal distributions. This would impose a large computational burden.

Instead we approximate $\pi_Z(\bftheta|\bfy)$ by a bivariate normal density $N_2({\bftheta}^*_y,\Sigma_y)$, 
  where ${\bftheta}^*_y=\argmax_{\bftheta} p_Z(y|\bftheta) \pi(\bftheta)$ and $\Sigma_y$ is the inverse of the Hessian matrix of $-\log(p_Z(y|\bftheta) \pi(\bftheta))$ (i.e.~of the negative log of the tractable auxiliary posterior) evaluated at ${\bftheta}^*_y$. In this manner, the auxiliary model $g(\bftheta|\bfs)$ is specified by this $N_2({\bftheta}^*_y,\Sigma_y)$ distribution, with $\bfs=({\bftheta}^*_y,\Sigma_y)^\top$, and the marginal distribution functions $\tilde{F}_{j,\bfs}(\cdot)$ are immediately available as univariate normal distribution functions. Calculation of ${\bftheta}^*_y$ and $\Sigma_y$ is very quick.

We simulate $n=10$ observations from the true model $Y = \sum^{10}_{\ell=1} X_{\ell}$, where $X_{\ell} \sim \mbox{LogNormal}(0, 1)$, to produce the observed dataset $\yobs$.
Algorithm \ref{alg:recalibration-aux} was then used to  generate $N=10,000$ approximate posterior samples.
For simplicity, we specified $h=\infty$ so that the weights $w^{(i)}=1/N$ were all equal. This provides a challenging scenario as we are then attempting to recalibrate all samples drawn from the prior to behave as approximate samples from $\pi(\bftheta|\sobs)$.

\begin{figure}[tb]
  \centering
  \subfloat[\footnotesize{Densities for $\bfy|\bftheta=(0, 1)^\top$}. \label{Density}]{\includegraphics[width=7cm,height=7cm,angle=-90]{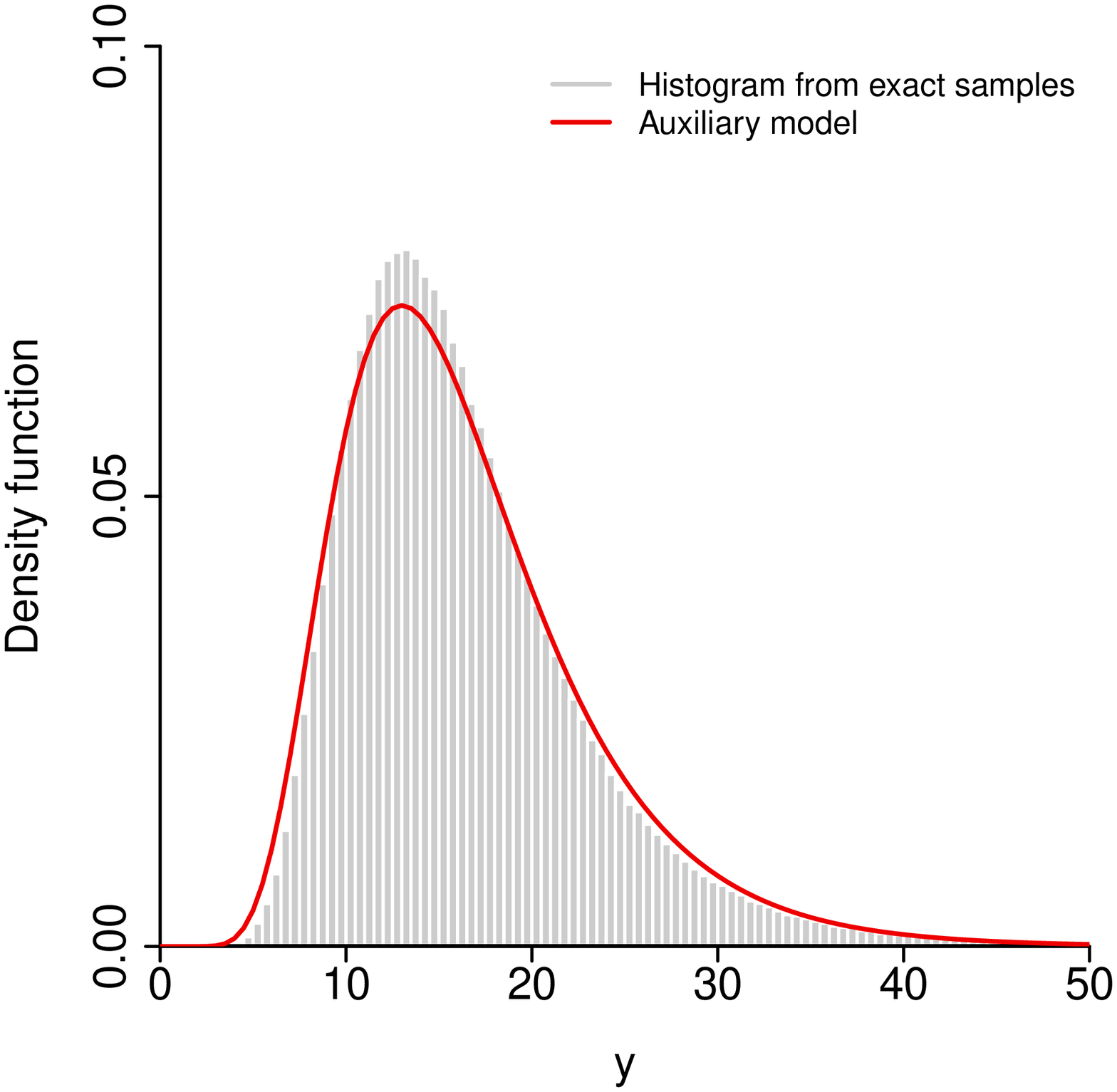}} 
  \subfloat[\footnotesize{Posterior density estimates}. \label{Posterior}]{\includegraphics[width=7cm,height=7cm,angle=-90]{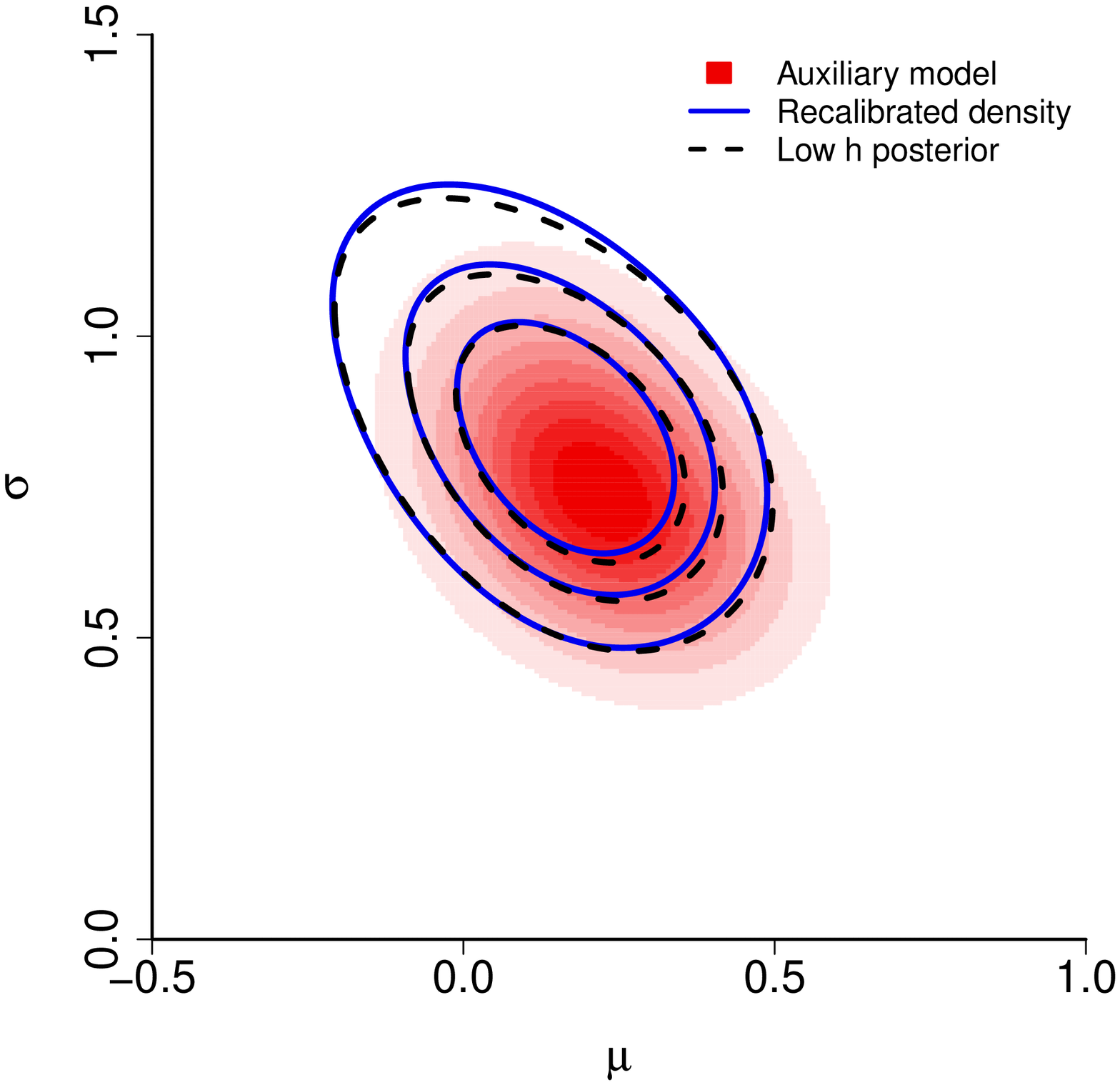}} \\
  \subfloat[\footnotesize{Realised samples from $G_{\bfy}(\bfp)$}. \label{pvalues}]{\includegraphics[width=7cm,height=7cm,angle=-90]{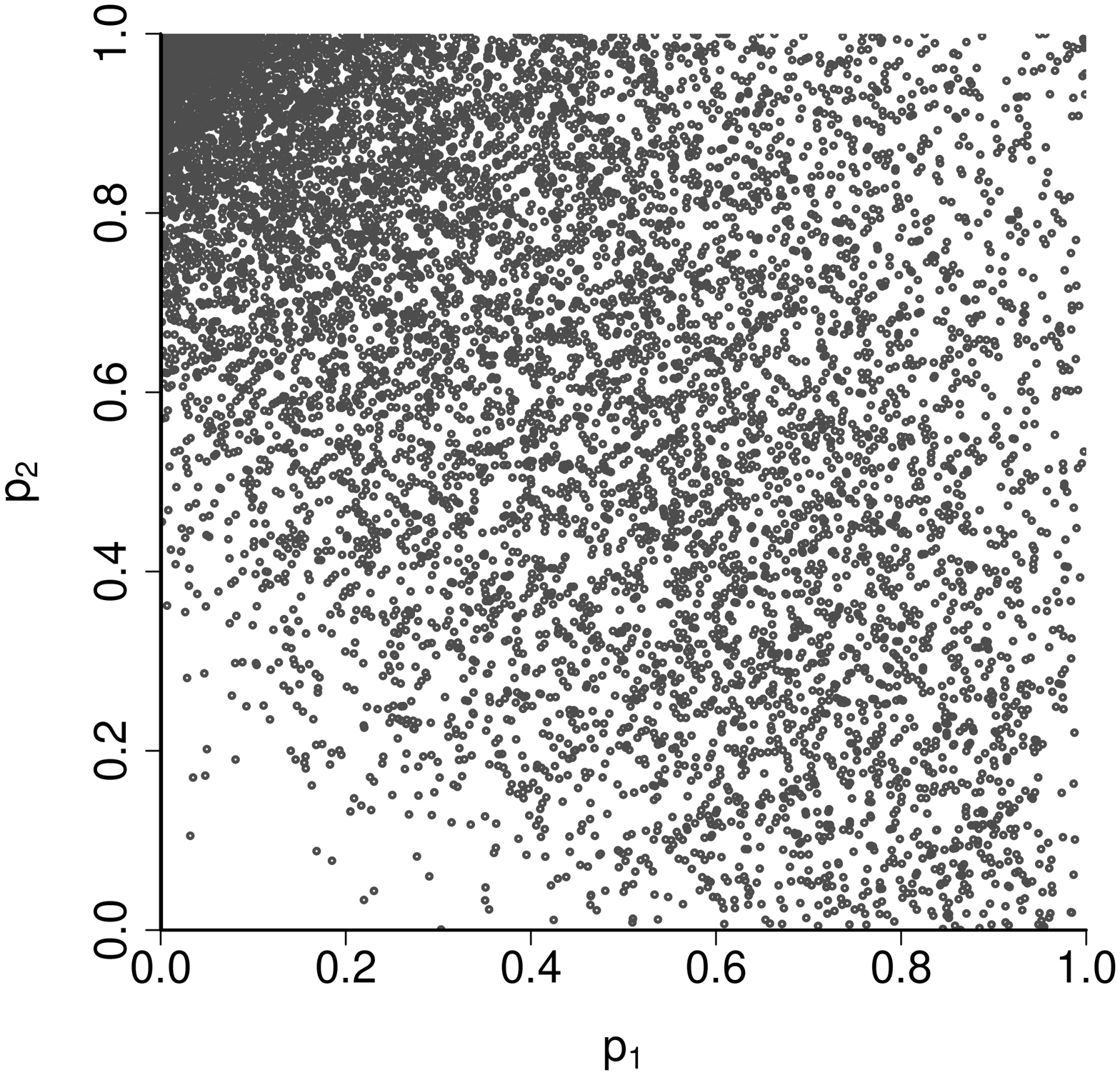}} 
  \subfloat[\footnotesize{Marginal distributions of $p_1$ and $p_2$}. \label{Histograms}]{\includegraphics[width=7cm,height=7cm,angle=-90]{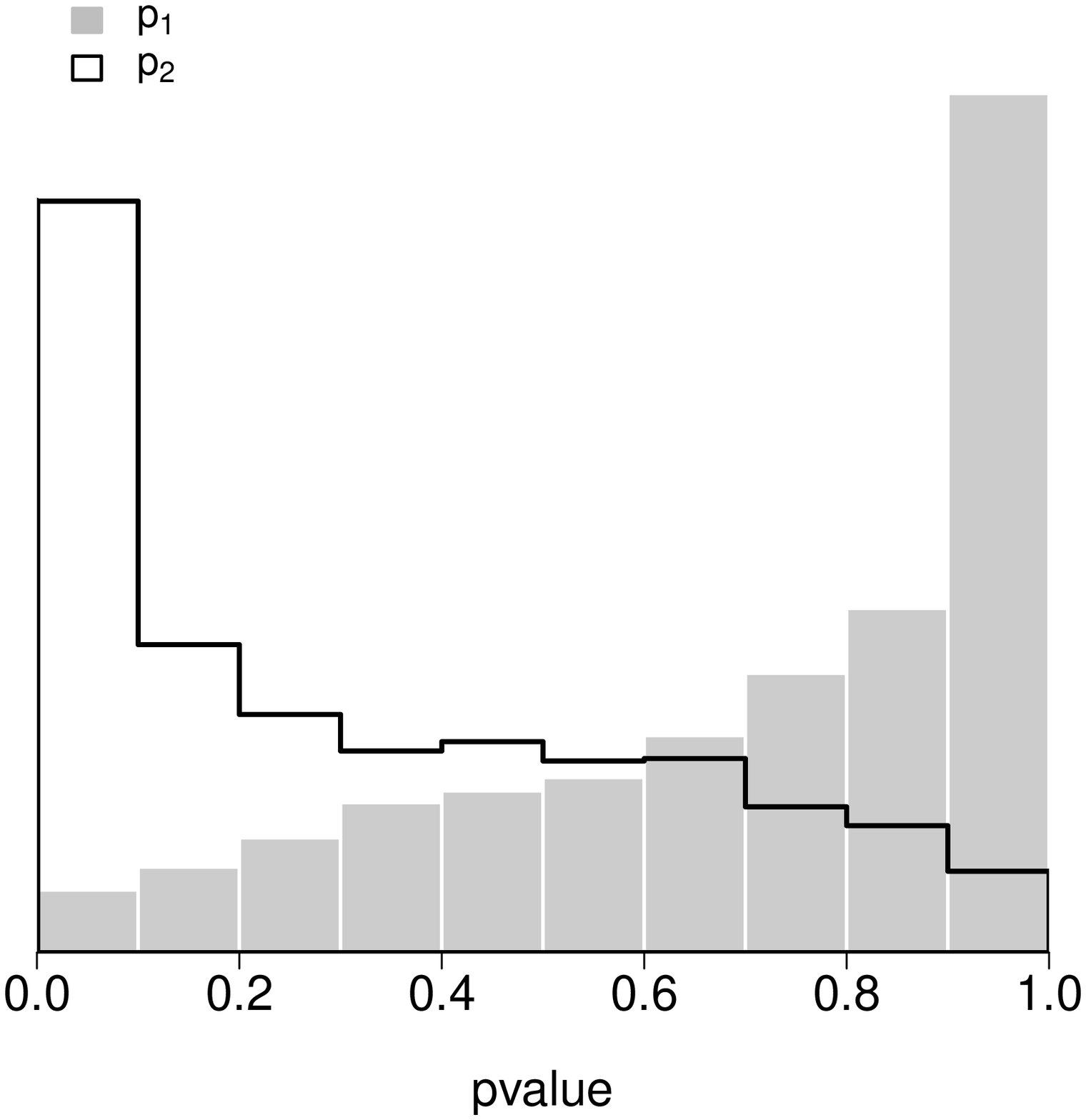}}
  \caption{\small Panel (a) compares the true density (histogram), $p_Y(\bfy|\bftheta=(0,1)^\top)$, with the corresponding Fenton-Wilkinson approximation $p_Z(\bfy|\bftheta=(0,1)^\top)$ (solid line).
Panel (b) compares kernel density estimates (KDE) of the approximate posterior resulting from: a low-$h$ ABC sampler (dashed line), the Fenton-Wilkinson auxiliary model (shading) and the recalibrated posterior (solid lines). Panels (c) and (d) respectively present the joint and marginal $\bfp=(p_1,p_2)^\top$ values obtained during recalibration.}
  \label{fig:example}
\end{figure}

Figure \ref{Density} compares the Fenton-Wilkinson lognormal density, $p_Z(\bfy|\bftheta)$, with the true density $p_Y(\bfy|\bftheta)$ at the true parameter values of 
$\bftheta=(0,1)^\top$. The lognormal density is clearly a reasonable match for the true density in this case, although it is slightly more diffuse. However the resulting posterior estimate (shading) is inaccurate, as illustrated in Figure \ref{Posterior}, compared to that obtained under a
 highly computational ABC rejection sampler (dashed lines) with the vector $\bfs=S(\bfy)={\bftheta}^*_y$ as summary statistics and with the kernel scale parameter $h$ reduced to a very low level. (The use of the MLE of a tractable approximation as summary statistics is a common approach.)
In contrast, the resulting recalibrated posterior approximation (solid lines) appears visually very close to the low-$h$ posterior.

A bivariate scatterplot and univariate marginal histograms of the $\bfp=(p_1,p_2)^\top$ values produced in the recalibration are shown in Figures \ref{pvalues} and \ref{Histograms}.
The non-uniformity of the marginal histograms suggests that the Fenton-Wilkinson method overestimates $\mu$ and underestimates $\sigma$ for this analysis, which is supported by the posterior density estimates in \ref{Posterior}.
In this case the recalibration procedure corrects these errors successfully.
In this analysis, the entire inference process took only a few seconds to complete on a desktop PC,
with the computational cost dominated by the optimization process involved in computing ${\bftheta}^*_y$.
In comparison, the cost of recalibration was negligible, as it only involved calculating $\bfp$ and quantiles from univariate normal distributions.

\subsection{A ``twisted normal'' model}
\label{sec:simulation}

In this analysis, we  investigate and quantify the effect of recalibration of standard ABC sampler output under various conditions. 
We consider the simple, deterministic data-generating model $Y = \theta_1 + \theta_2^2$, with $\bftheta=(\theta_1,\theta_2)^\top$, and suppose that $\theta_1$ and $\theta_2$ have independent $N(0, 1)$ priors. 
For a single observed data point $\yobs=y$, the resulting posterior mass is then concentrated on the set of points satisfying $\theta_1 = y-\theta_2^2$. 
For the below analysis we adopt $\yobs=1.$

\begin{figure}[tb]
  \centering
  \subfloat[\footnotesize{Effect of recalibration}. \label{mu}]{\includegraphics[width=7cm,height=7cm,angle=-90]{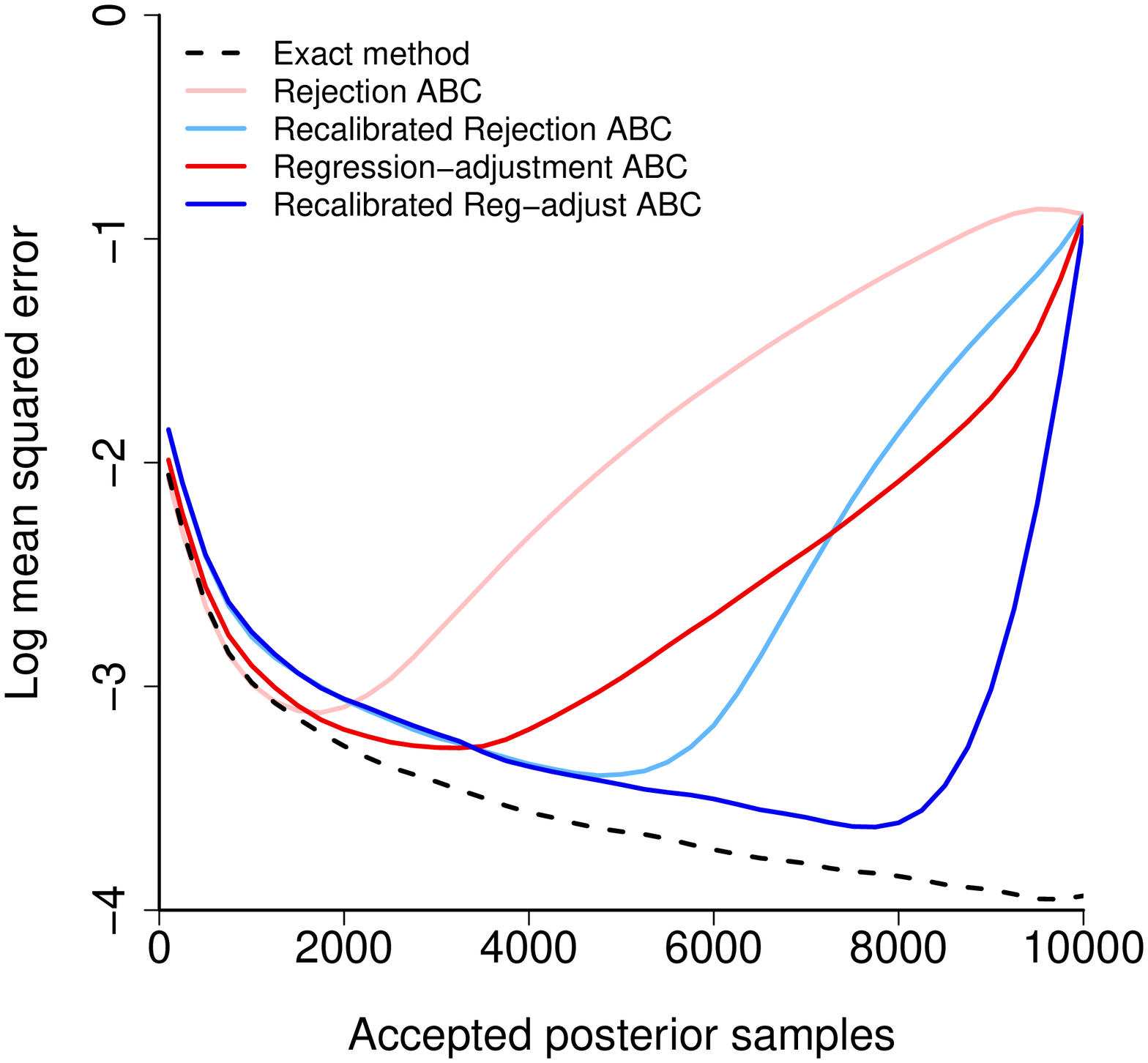}} 
  \subfloat[\footnotesize{Effect of correcting $\bfp$ values}. \label{mu_1}]{\includegraphics[width=7cm,height=7cm,angle=-90]{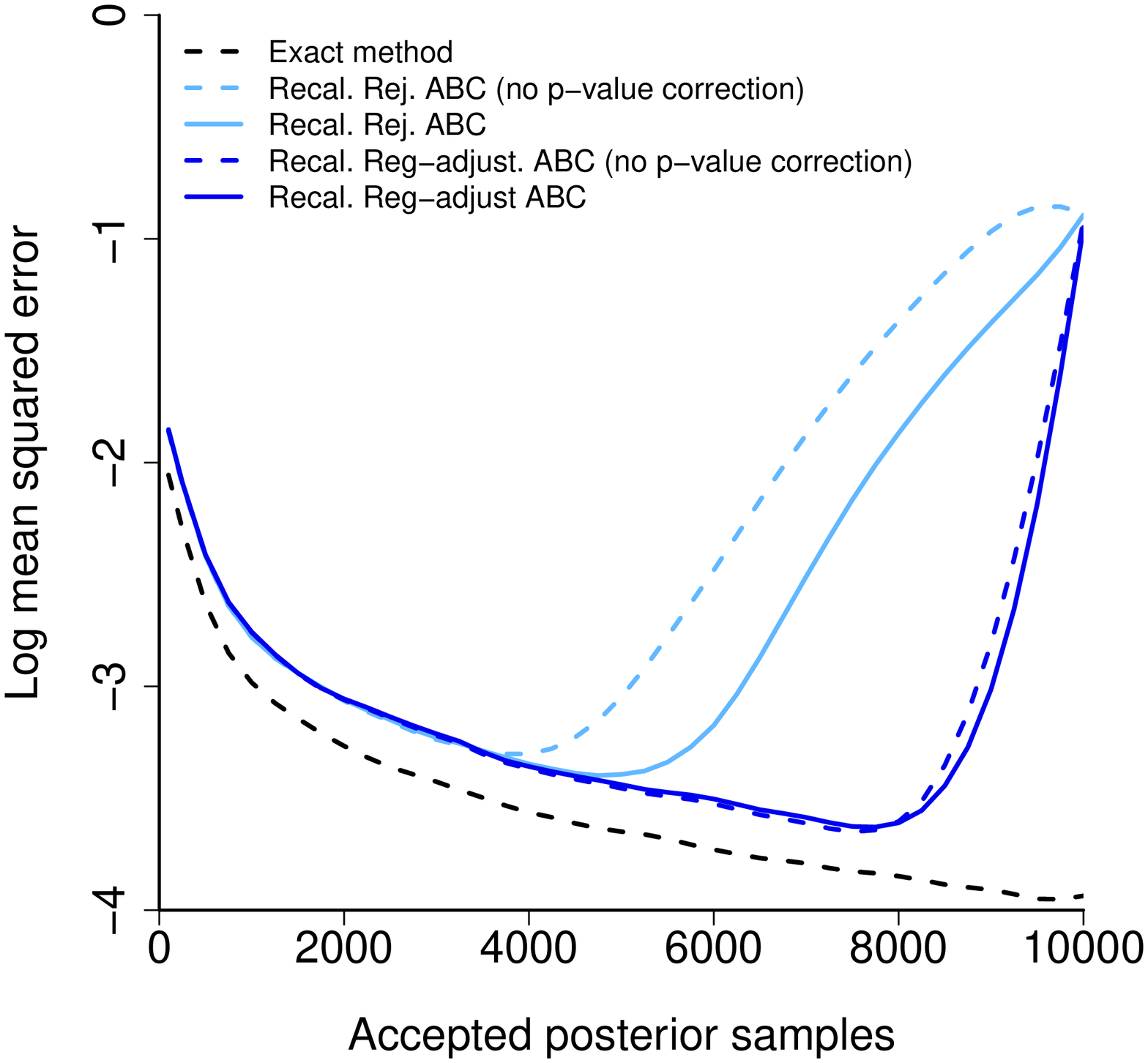}}
  \caption{\small Log mean squared error of different ABC methods when estimating $E(\theta_1-\theta_2|\theta_1+\theta_2^2=1)$, as a function of the number of posterior samples (out of $N=10,000$).  Panel (a)  compares rejection (red lines) and recalibrated (blue lines) ABC estimators. Darker and lighter lines respectively denote rejection and regression adjustment ABC. The dashed black line depicts the case when the samples were drawn from the exact posterior. Panel (b) contrasts log MSE for recalibrated ABC methods both with and without regression adjusted $\bfp$ values.}
  \label{fig:comparison}
\end{figure}

We follow Algorithm \ref{alg:recalibration}, and draw $N=10,000$ samples from the prior distribution, use the full dataset $\bfy$ (a single data point) as the summary statistic, and adopt the Epanechnikov kernel $K_h$, with $h$ determined by giving the 3,000 samples $\bftheta^{(i)}$ for which $\bfs^{(i)}$ is closest to $\sobs$ non-zero weights $w^{(i)}$ (e.g. \shortciteNP{biau+cg15}). The 30\% acceptance rate of the algorithm is approximately optimal for regression adjustment ABC in this analysis, in terms of producing the minimum mean square error (MSE) of a particular posterior functional (see below and Figure \ref{mu}).

\begin{figure}[tp]
  \centering
  \subfloat[\footnotesize{Regression-adjusted ABC}. \label{samples_reg}]{\includegraphics[width=5.6cm,height=5.6cm,angle=-90]{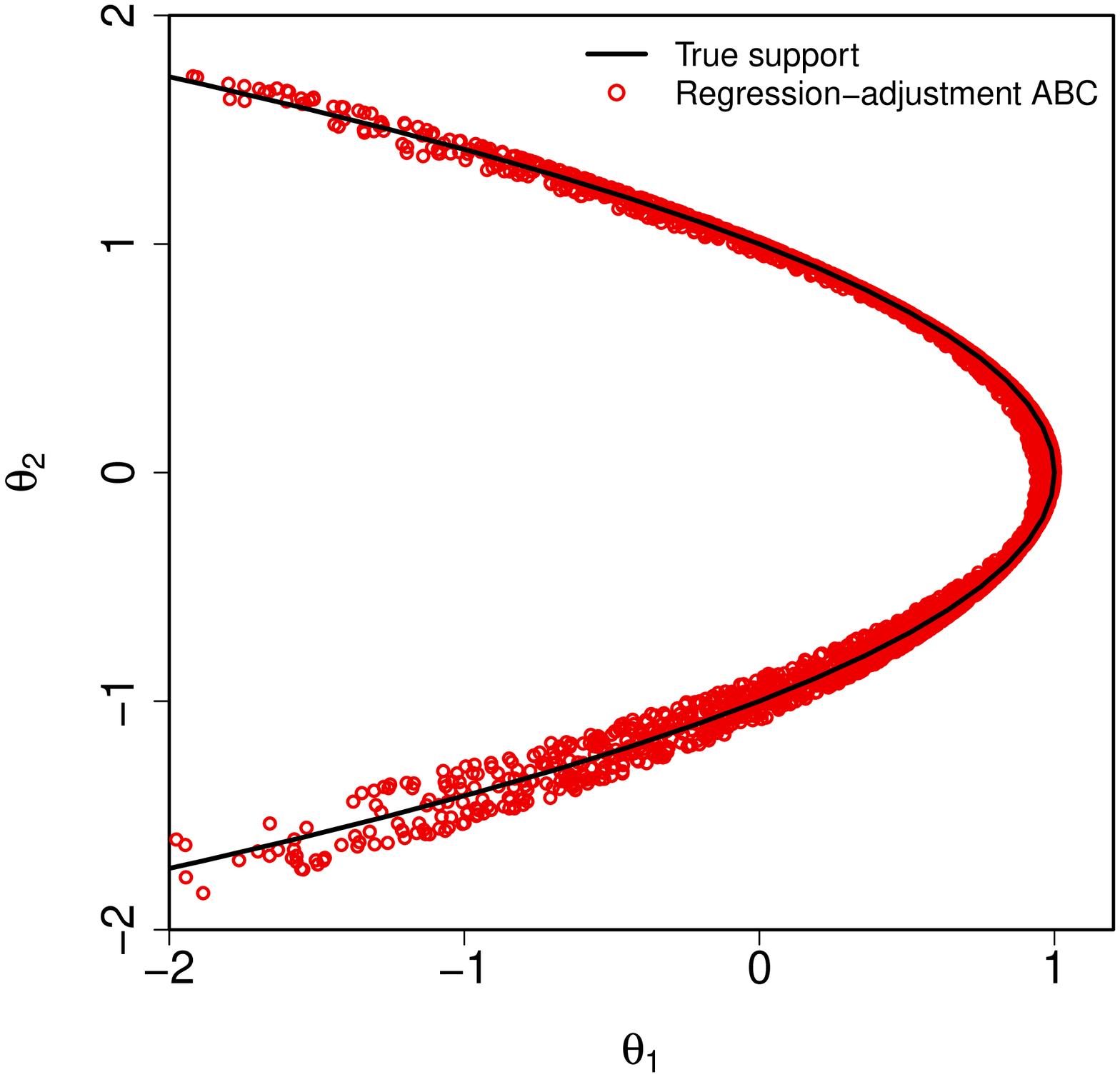}} 
    \subfloat[\footnotesize{Recalibrated posterior samples}. \label{samples_recal}]{\includegraphics[width=5.6cm,height=5.6cm,angle=-90]{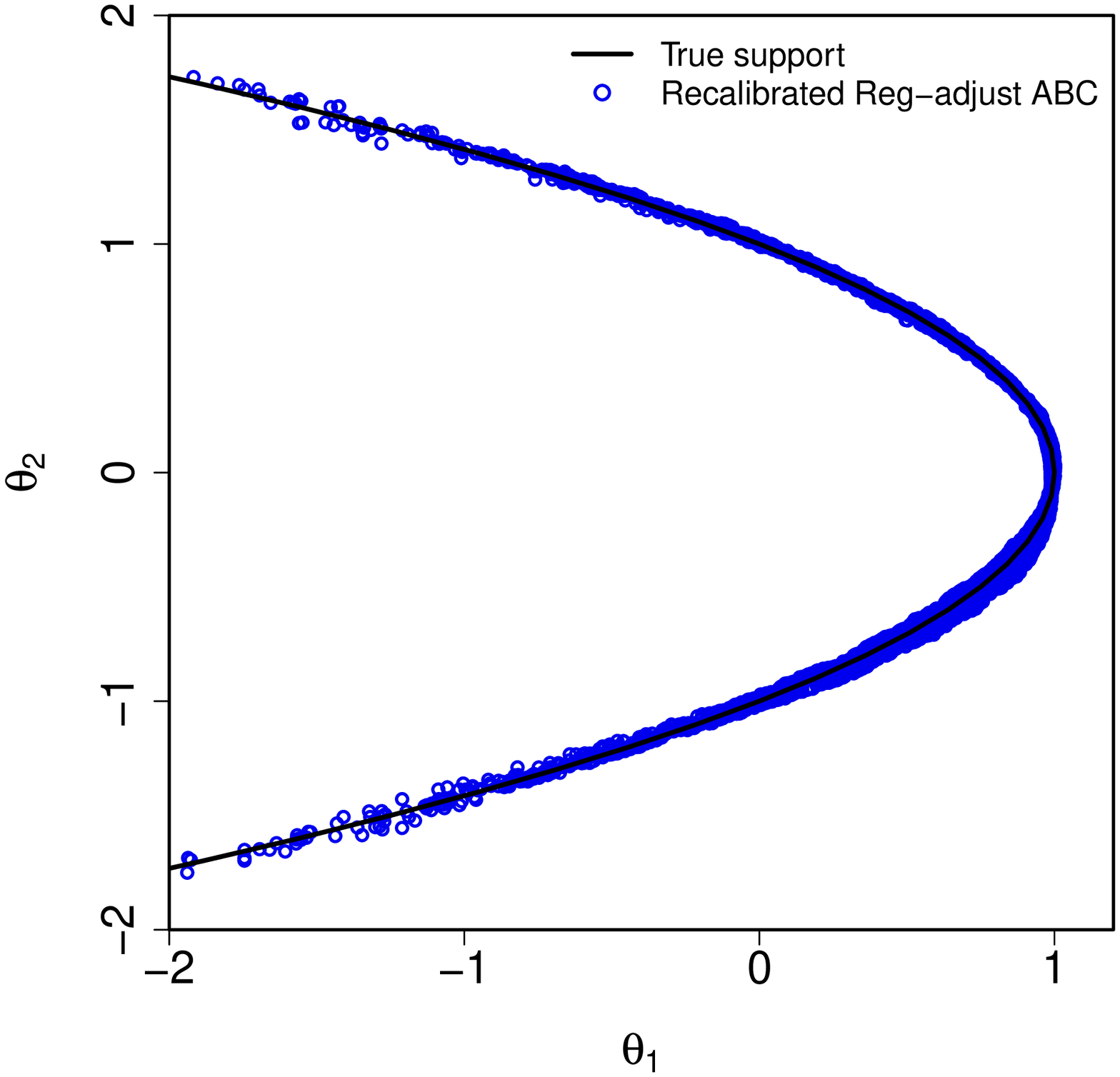}} 
  \subfloat[\footnotesize{Realised $\bfp$ values}. \label{pvalues.sim}]{\includegraphics[width=5.6cm,height=5.6cm,angle=-90]{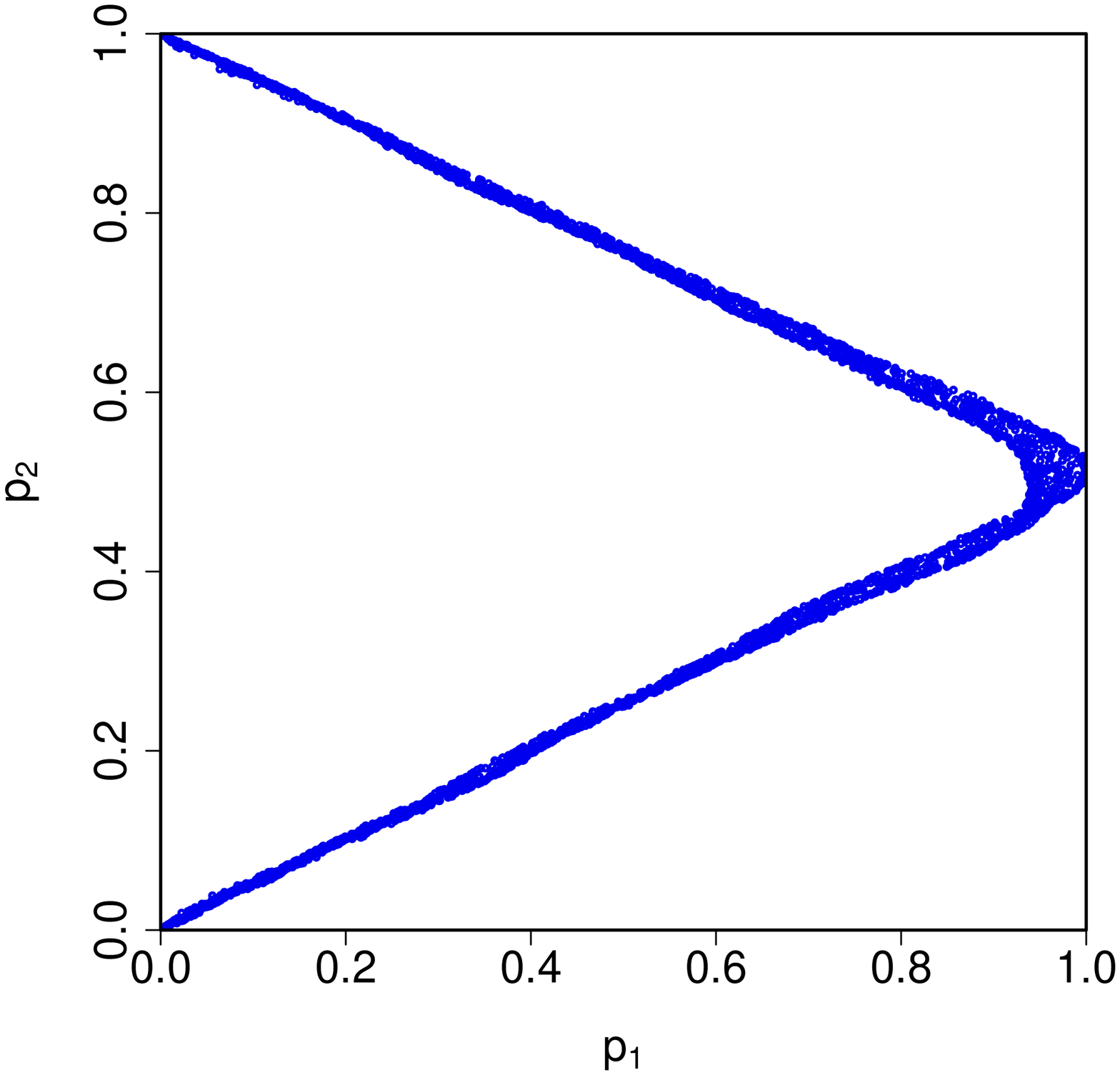}}
  \caption{\small Panel (a) illustrates 3,000 samples from posterior distribution estimates using regression adjusted ABC and panel (b) the same samples following recalibration. The grey line indicates the support of the true posterior. Panel (c) presents the corresponding realised $\bfp=(p_1,p_2)^\top$ values.}
  \label{fig:simulation1}
\end{figure}

Figure \ref{samples_reg} illustrates the regression-adjusted ABC samples in comparison to the support of the true posterior, shown by the solid line. Figure \ref{samples_recal} shows the same samples following recalibration, which includes the $\bfp$ value regression adjustment of Section \ref{sec:regadj}. 
Standard regression-adjustment ABC is easily able
to recover the twisted normal shape of the true posterior distribution, however the ABC approximation error is reflected by the extent of the samples lying far from the true posterior support (the solid line).
The recalibrated samples, while still having some deviation away from the true posterior support, visibly produce an improved posterior approximation.
This is particularly evident in the lower tail of the $\theta_2$ margin.

Figure \ref{pvalues.sim} shows the bivariate distribution of the realised $\bfp=(p_1,p_2)^\top$ values. Here, the univariate marginal distributions are almost uniform, indicating that the marginal posterior distributions of the regression-adjusted ABC posterior approximation are close to the true posterior marginal distributions  (c.f. Result 1 and \shortciteNP{Prangle2013}), while the striking dependence structure is a direct result of the form of $\pi(\bftheta|\sobs)$.

More qualitatively, we consider estimation of the posterior expectation $E(\theta_1-\theta_2|\theta_1+\theta_2^2=1)$ under each of four ABC posterior approximation procedures: standard rejection sampling ABC both with and without regression adjustment, and each of these with a subsequent recalibration adjustment (including a regression adjustment on the $\bfp$ values). This computation was repeated 1,000 times and for a range of Epanechnikov kernel scale parameter values $h$, resulting in between 100 and all 10,000 samples with non-zero weight $w^{(i)}>0$ being used for the computation. The log (base 10) mean squared error (MSE) over these 1,000 replicates  was recorded. The  conclusions of the below analysis were unchanged when other quantities of potential interest such as $P(\theta_1 > \theta_2|\theta_1+\theta_2^2=1)$ were considered.

Figure \ref{mu} displays the log of the MSE for each method as a function of the number of posterior samples (out of 10,000). The same quantity based on samples drawn from the exact posterior is illustrated by the dashed line. 
Each of the ABC based log MSE curves behave in a similar way as the number of posterior samples increases (i.e. as the kernel scale parameter $h$ increases). For small scale parameter values, the log MSE initially decreases as long as the quality of the posterior approximation for each method is high, with the decrease in log MSE achieved through an increase in the number of samples. That is, the high log MSE for low $h$ is primarily driven by Monte Carlo error. At some point, however, with increasing $h$ the quality of the posterior approximation deteriorates too much, and the log MSE increases due to bias in the posterior approximation.

However, the relative performance of each ABC method differs in its performance for low $h$, and  the point at which the bias in the posterior approximation begins to dominate the MSE. For low $h$ values standard rejection ABC (light red line) performs as well as the exact posterior distribution until around 1,500 samples. For low $h$, implementing any post-processing method only increases the Monte Carlo error, as these require the estimation of regression parameters and/or marginal distribution functions $\tilde{F}_{j,\bfs}(\cdot)$, with more overheads required for recalibration than for regression adjustment. 
For larger $h$, however, there is a clear benefit to post-processing, with the quality of the regression adjusted posterior approximation (dark red line) meaning that it can reach a lower log MSE for an $h$ equivalent to around 3,000 samples. The recalibrated posterior approximations perform even more efficiently, with the recalibrated regression-adjusted ABC posterior the most efficient of all, achieving their optimum log MSE values at around 5,000 and 8,000 samples. 
In fact, the minimum MSE obtained by recalibration (recalibrated  regression-adjusted ABC) was 0.0002, which is a sizeable reduction from its uncalibrated counterpart of 0.0005 (regression-adjusted ABC) -- especially taking into account the theoretical minimum, 0.0001, obtained by exact calculations.

Figure \ref{mu_1} presents the same information as Figure \ref{mu} but comparing the recalibration adjusted methods both with and without regression adjusted $\bfp$ values (Section \ref{sec:regadj}).
 In this case, adjusting the $\bfp$ values  clearly improves recalibrated rejection ABC, but  recalibrated regression-adjustment ABC is only improved to a small extent.
 This primarily occurs as the linear regression model assumptions are not reasonable in this region.

In the above analysis, for ease of presentation, the same acceptance rate adopted in steps 1.4 and 2.1 of Algorithm \ref{alg:recalibration} was used when computing the marginal estimates $\tilde{F}_{j,\bfs}(\cdot)$ in step  2.2.
However, it could be computationally more efficient to use different rates for each step, such as using 30\% of the synthetic samples to recalibrate a regression-adjustment ABC based on an acceptance rate of 10\%.

\section{Application: Estimation in Stereological extremes}
\label{sec:application}

During the production of a steel block, endogenous or exogenous chemical compounds are unavoidably embedded into the final product. Known as \emph{inclusions}, these foreign substances affect the toughness, corrosion resistance and other features of the steel. 
The size of the largest inclusions, which cannot be directly observed, are particularly influential to the overall quality. Therefore, interest lies in an extreme value problem in which  inference is required on the distribution of the largest inclusion sizes based on the inclusions observed in a two-dimensional planar slice through the block. Each observed cross-sectional inclusion size in $\yobs = (y_{\mathrm{obs}, 1}, \ldots, y_{\mathrm{obs}, n})^\top$ is related to an unknown inclusion size $V_i>y_{\mathrm{obs}, i}$ in 3-dimensional space. The number of inclusions in the sample is random, and, for any given $i$, the probability of observing $y_{\mathrm{obs}, i}$ depends on $V_i$ -- larger inclusions are more likely to intersect the planar slice.

To make inference in this stereological context,
 it is commonly assumed that the inclusion centres follow a homogeneous Poisson process with rate $\lambda$, and that inclusion sizes are mutually independent and independent of inclusion location. These assumptions are widely regarded as reasonable. When it comes to the shape of the inclusions, however, different formulations have been studied. 
\citeN{Anderson2002} assumed that inclusions were spherical, with ``size'' being characterized by the inclusion's diameter $V$. Subsequently \shortciteN{Bortot2007} 
considered randomly oriented ellipsoidal shapes, where $y_{\mathrm{obs}, i}$ then refers to the largest principal diameter of the $i$th observed ellipse and $V_i$ the  largest diameter of the corresponding ellipsoid. 
In both spherical and ellipsoidal constructions, a generalized Pareto distribution (GPD) is assigned to $V|V>v_0$, where $v_0$ is an appropriate threshold. The distribution function is given by
\[
P(V \leq v | V>v_0) = 1-\left[ 1+\frac{\xi(v-v_0)}{\sigma} \right]_+^{-1/\xi},
\]
where $[a]_+=\max\{0, a\}$, $v>v_0$, and $\sigma>0$ and $-\infty < \xi < \infty$ are scale and shape parameters. To fully specify the model, \shortciteN{Bortot2007} also assumed that the two non-leading principal diameters of a given ellipsoid are defined as $V_1=U_1 V$ and $V_2=U_2 V$, where $U_1$ and $U_2$ are independent standard uniform variables. 

\citeN{Anderson2002} derived an exact MCMC sampler for the posterior distribution of their spherical model. However, the likelihood induced by the more plausible ellipsoidal model is computationally intractable, which motivated \shortciteN{Bortot2007} to use ABC methods for inference on $\bftheta=(\lambda, \sigma, \xi)^\top$.
\citeN{Erhardt} conducted a simulation study to investigate the performance of different ABC implementations in this context,
demonstrating that regression-adjustment substantially improved the accuracy of rejection ABC.

They adopted a uniform prior distribution for $\bftheta$, restricted to a region that comfortably enveloped the effective support of the posterior distribution.   
In addition, they adopted the summary statistics
\begin{equation}
\label{ss.Erhardt} S(\bfy)=(n', q_{0.5}(\bfy), q_{0.7}(\bfy), q_{0.9}(\bfy), q_{0.95}(\bfy), q_{0.99}(\bfy), q_{1}(\bfy))^\top,
\end{equation}
where $q_a(\bfy)$ denotes the $a$-th quantile of $\bfy$, and $n'$ is the (random) number of observations in $\bfy$. Their ABC analyses were performed using the best 2,000 out of $N=2$ million generated samples $\{(\bftheta^{(i)},\bfs^{(i)})\}_{i=1}^N$. 

\begin{figure}[tp]
  \centering
    \subfloat[\footnotesize{Marginal distribution of $p_\xi$}. \label{app.hist.spherical}]{\includegraphics[width=7cm,height=7cm,angle=-90]{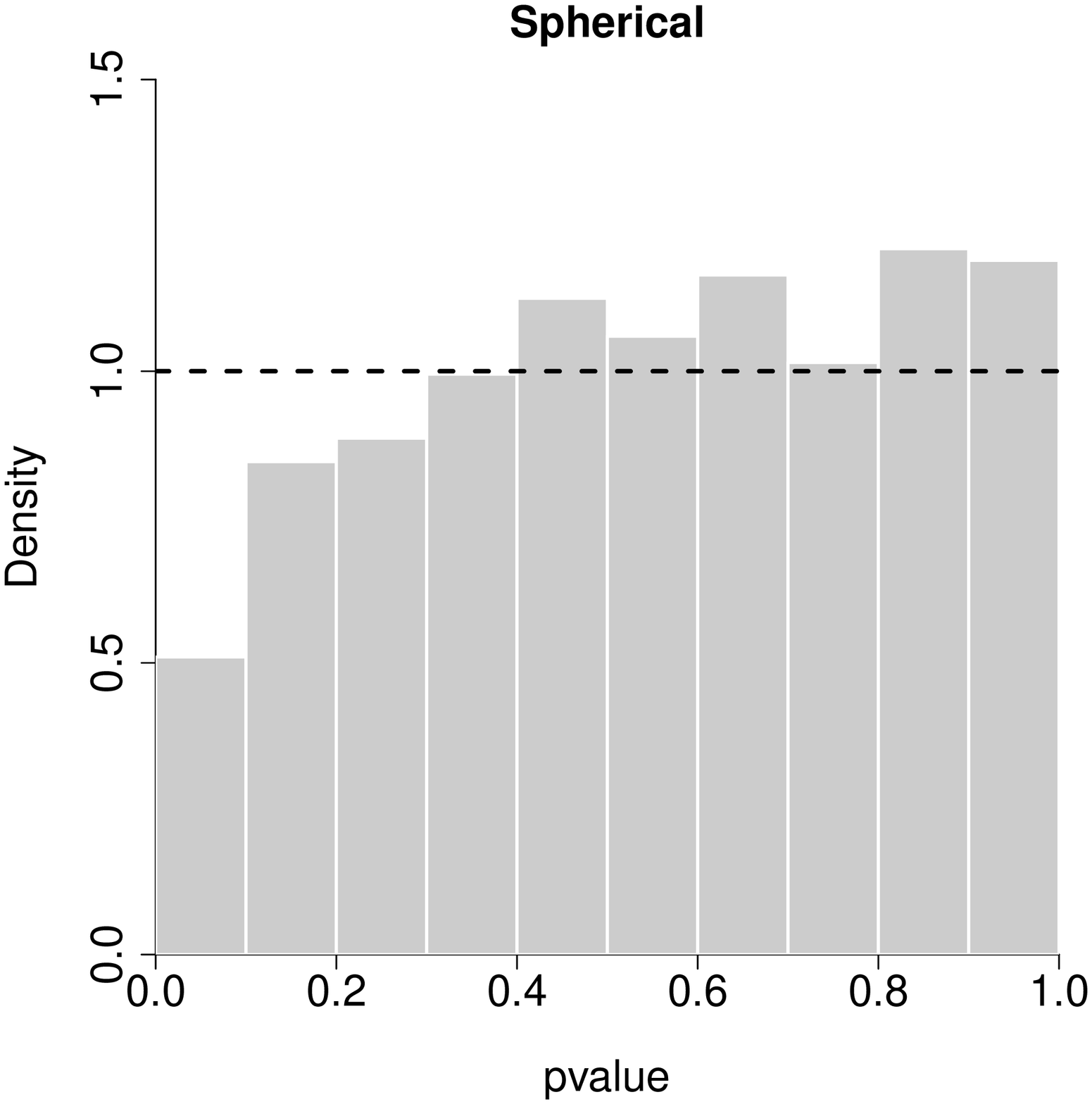}} 
  \subfloat[\footnotesize{Marginal distributionsof $p_\xi$}. \label{app.hist.elliptical}]{\includegraphics[width=7cm,height=7cm,angle=-90]{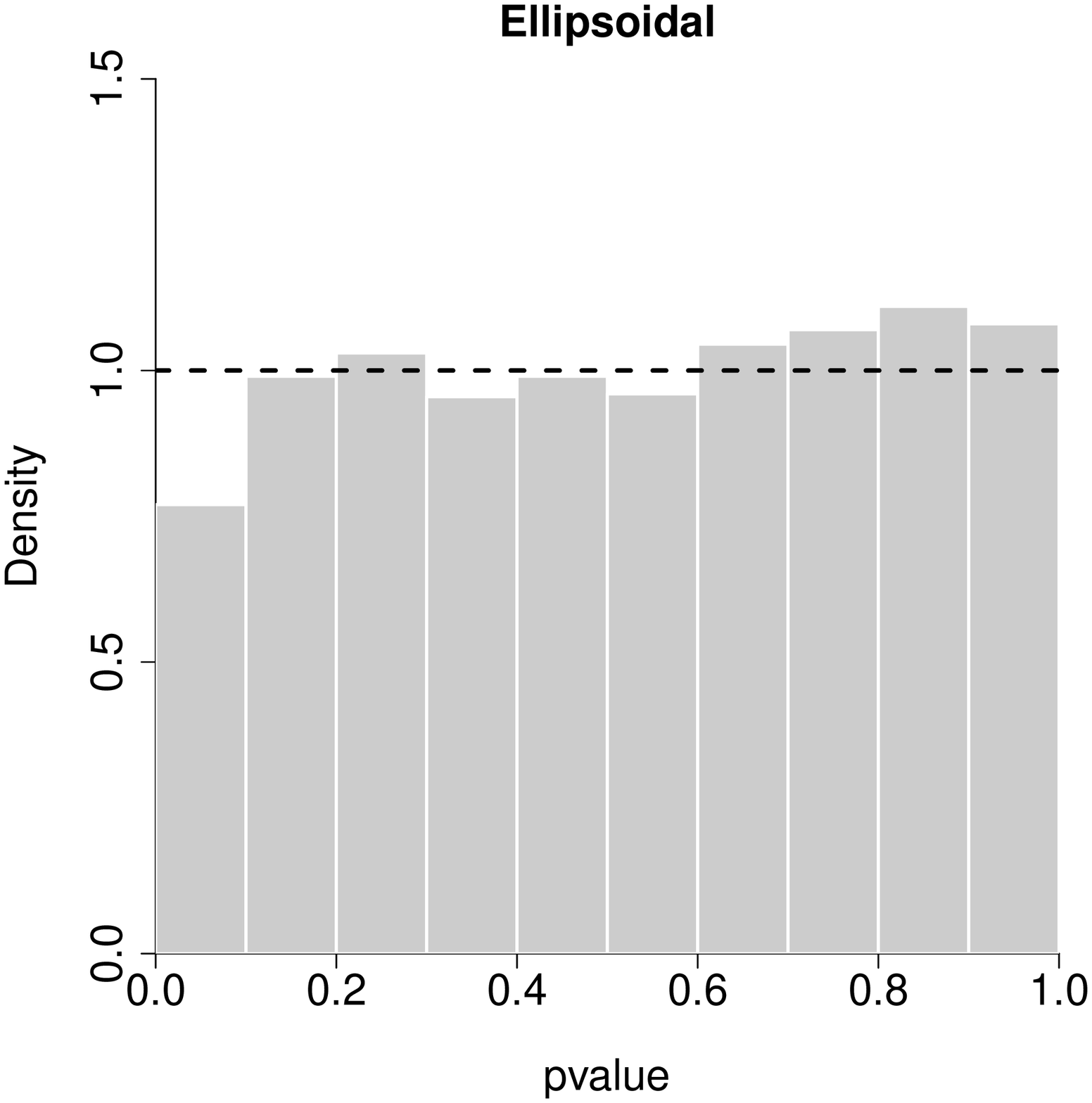}} \\
  \subfloat[\footnotesize{Estimated posterior of $\xi$}. \label{app.dens.spherical}]{\includegraphics[width=7cm,height=7cm,angle=-90]{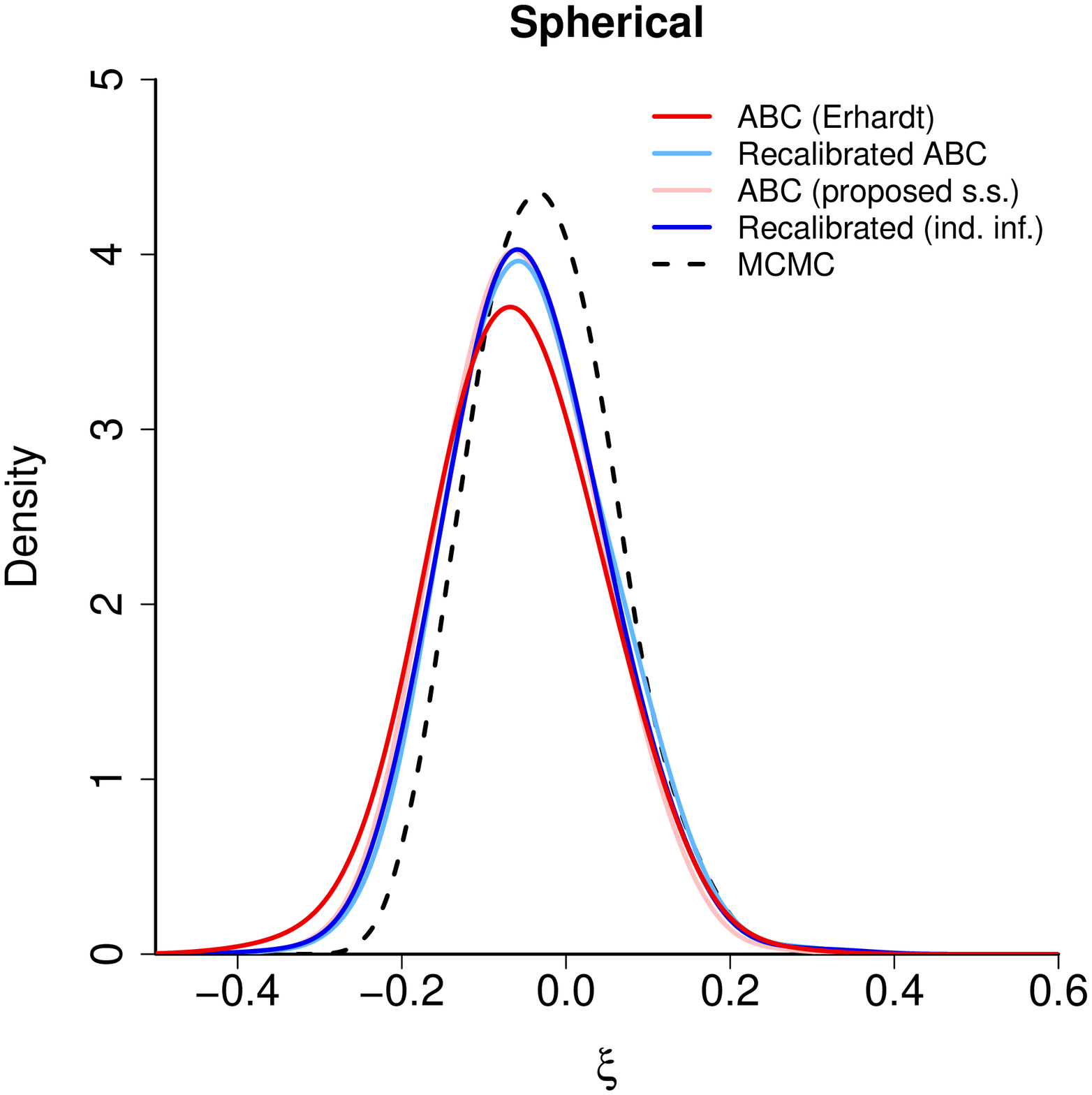}} 
  \subfloat[\footnotesize{Estimated posterior of $\xi$}. \label{app.dens.elliptical}]{\includegraphics[width=7cm,height=7cm,angle=-90]{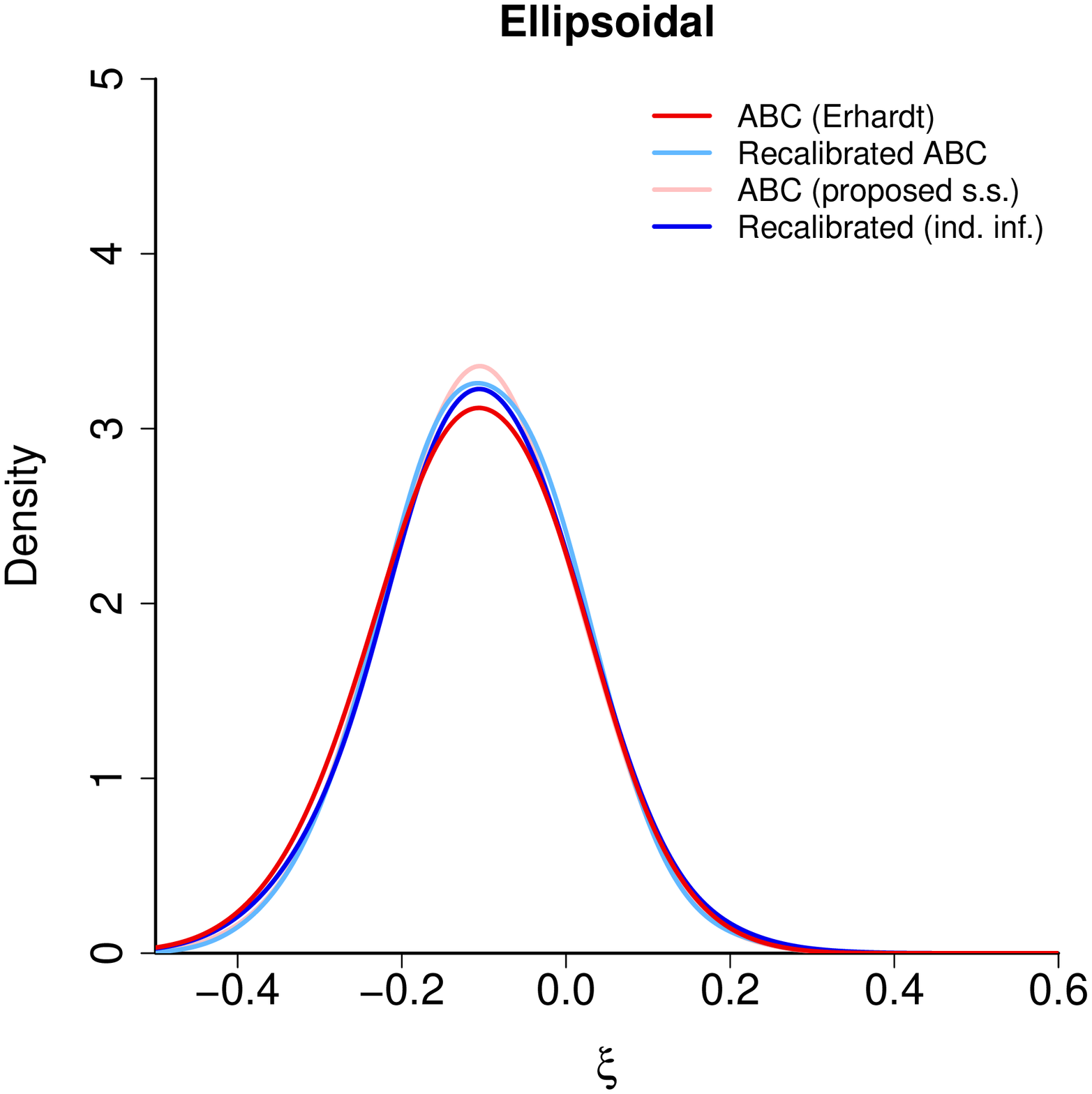}}
  \caption{\small Panels (a) and (b) show for the spherical and elliptical cases, respectively, the realised $\bfp$ values, $p_\xi$, associated with the recalibration of regression-adjustment ABC using the  summary statistics \eqref{ss.Erhardt}.  Panels (c) and (d) compare the marginal posterior densities for $\xi$  estimated by different methods and summary statistics.
}
  \label{fig:application}
\end{figure}

With these same settings, we revisit the analysis in \citeN{Erhardt}, using Algorithm \ref{alg:recalibration} to generate recalibrated samples from the regression-adjustment ABC posterior approximation. We focus our attention on the shape parameter $\xi$ as it determines the tail behaviour of extreme value models.
We also investigate recalibrating a computationally cheaper auxiliary method using Algorithm \ref{alg:recalibration-aux}, similar to that implemented in Section \ref{sec:example}. In the stereological context, intractability arises from the impossibility to measure the diameters $V_i$. We therefore use a tractable, but misspecified, auxiliary model which assumes that the observable diameters, $\bfy|\bfy>v_0$, follow a GPD with parameters $\sigma'$ and $\xi'$. A new set of summary statistics may then be defined as
\[
S'(\bfy)=(n', \tilde{\sigma}(\bfy), \tilde{\xi}(\bfy))^\top,
\]
where $\tilde{\sigma}(\bfy)$ and $\tilde{\xi}(\bfy)$ are the MLEs of this auxiliary model. Although highly informative, $S'(\bfy)$ is not itself an estimator for $\bftheta$. So for each simulated dataset $\bfs'^{(i)}$, we follow \citeN{fearnhead2012} and estimate $\bftheta^{(i)}$ by 
${\bftheta}^{+(i)}$, where 
${\bftheta}^{+(i)}_j=E(\bftheta^{(i)}_j|\bfs'^{(i)}_j)$,
using univariate (splines) smoothers fitted to $\{(\bftheta_j^{(i)},\bfs_j'^{(i)})\}_{i=1}^N$ for $j=1,\ldots,d$, using the default settings of the {\tt smooth.spline} function in {\tt R}. Finally, we define a Gaussian auxiliary marginal estimator
as $\tilde{F}_{j,\bfs'^{(i)}}(\theta_j)=\Phi(\theta_j; {\bftheta}_j^{+(i)}, \hat{\sigma}_j)$, where $\hat{\sigma}_j$ is the standard deviation of the spline residuals for parameter $j$.

Figures \ref{app.hist.spherical} and \ref{app.hist.elliptical} show the distribution of the marginal $\bfp$ values for $\xi$, $p_\xi$, obtained when recalibrating the best ABC estimator considered in \citeN{Erhardt} -- namely, regression-adjustment ABC with summary statistics given by (\ref{ss.Erhardt}). The left-skew of both plots indicates that this regression-adjustment ABC tends to underestimate $\xi$.  A  Kolmogorov-Smirnov test rejects the hypothesis that the $p_\xi$ samples are from a $U(0,1)$ distribution (with $p$-values of $7 \times 10^{-11}$ and $0.02$ for the spherical and ellipsoidal cases respectively). 

Figures \ref{app.dens.spherical} and \ref{app.dens.elliptical} compare the marginal posterior density estimates for $\xi$ using regression adjusted ABC (red line) and its recalibration (light blue line), with the posterior estimates using regression adjusted ABC using the summary statistics $S'(\bfy)$ (pink line), and the recalibration of the Gaussian auxiliary estimator (dark blue line).
Also shown for the spherical model (dashed line) is the exact posterior obtained from the MCMC sampler of \citeN{Anderson2002}, although this is based on a partially-conjugate prior specification defined on a reparameterised space, and so this targets a different posterior to the ABC algorithms. Accordingly, a perfect correspondence between the exact posterior and the ABC methods should not be expected.

For the spherical model (Figure \ref{app.dens.spherical}) the underestimation of $\xi$ reflected in the $p_\xi$ values using the summary statistics (\ref{ss.Erhardt}) is visibly evident, and this is corrected under recalibration.  For the ellipsoidal case, the initial bias in $\xi$ was so mild that recalibration has barely affected the posterior estimate.  For both spherical and ellipsoidal models, standard regression-adjusted ABC with the new summary statistics $S'(\bfy)$ has performed as well as the recalibration of the Gaussian auxiliary estimator, with both densities appearing indistinguishable from the recalibrated standard ABC analysis. That these density estimates all lie in the same place strongly suggests that these are all good approximations to the true posterior in this case (with the uniform prior specification). It also suggests that the indirect inference-based summary statistics $S'(\bfy)$ are highly informative for these models. Overall, either adoption of $S'(\bfy$) or any method of recalibration produces a more accurate posterior approximation than the analysis performed in \citeN{Erhardt}.

\section{Discussion}
\label{sec:discussion}

This article introduces a recalibration procedure to post-process output from approximate Bayesian methods, in particular ABC techniques, based on the ideas in \shortciteN{Prangle2013}. Recalibration can improve the quality of an approximation of the posterior distribution by ensuring that the adjusted posterior estimate approximately satisfies the coverage property.
This means that errors and biases induced by adopting various posterior approximations, such as the standard ABC posterior approximation or auxiliary model approximations, can be (approximately) corrected. Indeed, this may then be exploited so that the most computationally efficient approximate posterior can be adopted, which is not necessarily standard ABC, in the knowledge that a good adjustment is available to correct model mis-specification.

Accordingly, in Section \ref{sec:example} the error induced by the incorrect assumption that a sum of log-normal distributions follows a log-normal distribution was substantially reduced by recalibration. 
Section \ref{sec:simulation} illustrated that recalibration can serve as a non-parametric alternative to regression-adjustment ABC (when an appropriate regression model is not available), or as an additional layer of post-processing to correct the biases of the regression-adjustment itself. 
In the stereological extremes analysis in Section \ref{sec:application}, using recalibration to correct a small bias in the results obtained by \citeN{Erhardt}, along with a more detailed investigation, provided a reassurance that more substantial errors have not been incurred in this analysis.

Recalibration does come with some computational cost, which may or may not be worthwhile, depending on a number of factors. An obvious practical requirement is that the auxiliary method used to construct the univariate marginal distributions $\tilde{F}_{j,\bfs}(\cdot)$ needs to be fast, or the computational overheads involved in recalibration will dominate those of the original analysis. 
Recalibration is also particularly appealing when simulation of datasets $\bfy\sim p(\bfy|\bftheta)$ under the model is computationally expensive. For instance, in the stereological extremes analysis of Section \ref{sec:application}, the recalibration stage of Algorithm \ref{alg:recalibration-aux} required no more than 10\% of the total computational time -- a modest computational cost for this analysis.

As with standard ABC methods, the best choice of kernel scale parameter $h$ is generally a non-trivial task. In principle, this choice is based on a balancing of Monte Carlo variation and the intrinsic error arising from assuming that $G_{\bfs}(\bfp)$ is nearly independent from $\bfs$  in the neighborhood of $\sobs$, as visualised in Figure \ref{fig:comparison}.
Further, as observed by \shortciteN{Prangle2013}, marginal uniform distributions for the realised $\bfp$ values are possible from distributions other than the true posterior distribution. In particular, if ABC or the auxiliary method returns the prior distribution $\pi(\bftheta)$ as the approximate posterior (see also the noisy ABC of \citeNP{fearnhead2012}), then as the prior automatically satisfies coverage \shortcite{Prangle2013}, recalibration post-processing will have no power to make a correction.

We have presented recalibration as a post-processing method for ABC and indirect inference based procedures. However, it may conceivably also be used for other methods for approximating posterior distributions, including variational methods and expectation propagation techniques.
%
An implementation of Algorithm \ref{alg:recalibration} is available in the {\tt abctools} {\tt R} package.

\subsection*{Acknowledgements}

 GSR is funded by the CAPES Foundation via the Science Without Borders program (BEX 0974/13-7).
 SAS is supported by the Australia Research Council through the Discovery Project Scheme (DP1092805), and the Australian Centre of Excellence for Mathematical and Statistical Frontiers in Big Data, Big Models, New Insights (ACEMS, CE140100049).
 DP was supported by
a Richard Rado Fellowship from the University of Reading during some of
this project.
 This research includes computations using the Linux computational cluster Katana supported by the Faculty of Science, UNSW Australia.

\bibliographystyle{chicago}
\bibliography{Recalibration_ABC}

\end{document}